\documentclass[lettersize,journal]{IEEEtran}
\usepackage{amsmath,amsfonts}
\usepackage{algorithmic}
\usepackage[ruled,linesnumbered]{algorithm2e}
\usepackage{array}
\usepackage[caption=false,font=footnotesize,labelfont=rm,textfont=rm]{subfig}
\usepackage{textcomp}
\usepackage{stfloats}
\usepackage{url}
\usepackage{verbatim}
\usepackage{multirow}
\usepackage{graphicx}
\usepackage{cite}
\usepackage{makecell}
\usepackage{enumitem}
\usepackage{graphicx}
\usepackage{float}
\usepackage{subfig}
\usepackage{lineno,hyperref}
\usepackage{booktabs}
\usepackage{xcolor}
\begin{document}

\title{TF4CTR: Twin Focus Framework for CTR Prediction via Adaptive Sample Differentiation}

\author{Honghao Li†, Qiuze Ru†, Yiwen Zhang*, Yi Zhang, Lei Sang, and Yun Yang, \textit{Senior Member, IEEE}

\IEEEcompsocitemizethanks{\IEEEcompsocthanksitem Honghao Li, Qiuze Ru, Yiwen Zhang, Yi Zhang, and Lei Sang are with the School of Computer Science and Technology, Anhui University 230601,
Hefei, Anhui, China (e-mail: salmon1802li@gmail.com, ruqiuze@stu.ahu.edu.cn, zhangyiwen@ahu.edu.cn, zhangyi.ahu@gmail.com, and sanglei@ahu.edu.cn).
\IEEEcompsocthanksitem Yun Yang is with Swinburne University of Technology, Hawthorn, Melbourne, Australia, VIC 3122. (e-mail:  yyang@swin.edu.au).
}
\thanks{†Equal contribution.}
\thanks{*Corresponding author.}}

\markboth{Journal of \LaTeX\ Class Files,~Vol.~x, No.~x, August~xxxx}%
{Shell \MakeLowercase{\textit{et al.}}: TF4CTR: Twin Focus Framework for CTR Prediction via Adaptive Sample Differentiation}


\maketitle
\begin{abstract}
Effective feature interaction modeling is critical for enhancing the accuracy of click-through rate (CTR) prediction in industrial recommender systems. Most of the current deep CTR models resort to building complex network architectures to better capture intricate feature interactions or user behaviors. However, we identify two limitations in these models: (1) the samples given to the model are undifferentiated, which may lead the model to learn a larger number of easy samples in a single-minded manner while ignoring a smaller number of hard samples, thus reducing the model's generalization ability; (2) differentiated feature interaction encoders are designed to capture different interactions information but receive consistent supervision signals, thereby limiting the effectiveness of the encoder. To bridge the identified gaps, this paper introduces a novel CTR prediction framework by integrating the plug-and-play \textit{Twin Focus (TF) Loss}, \textit{Sample Selection Embedding Module (SSEM)}, and \textit{Dynamic Fusion Module (DFM)}, named the Twin Focus Framework for CTR (TF4CTR). Specifically, the framework employs the SSEM at the bottom of the model to differentiate between samples, thereby assigning a more suitable encoder for each sample. Meanwhile, the TF Loss provides tailored supervision signals to both simple and complex encoders. Moreover, the DFM dynamically fuses the feature interaction information captured by the encoders, resulting in more accurate predictions. Experiments on five real-world datasets confirm the effectiveness and compatibility of the framework, demonstrating its capacity to enhance various representative baselines in a model-agnostic manner. To facilitate reproducible research, our open-sourced code and detailed running logs will be made available at: \url{https://github.com/salmon1802/TF4CTR}.
\end{abstract}


\section{Introduction}
\IEEEPARstart{C}{lick-through} Rate (CTR) prediction is crucial for industrial recommender systems \cite{openbenchmark, Bars, CVGA, DVGRL}, leveraging user profiles, item attributes, and context to predict user-item interactions. Accurate CTR predictions significantly influence system profits \cite{dcnv2, widedeep, EDCN, deepfm}, while also improving user satisfaction and retention through better recognition of user interests, enhancing the overall experience.

\begin{figure}[t]
    \begin{minipage}[t]{1\linewidth}
        \centering
        \includegraphics[width=\textwidth]{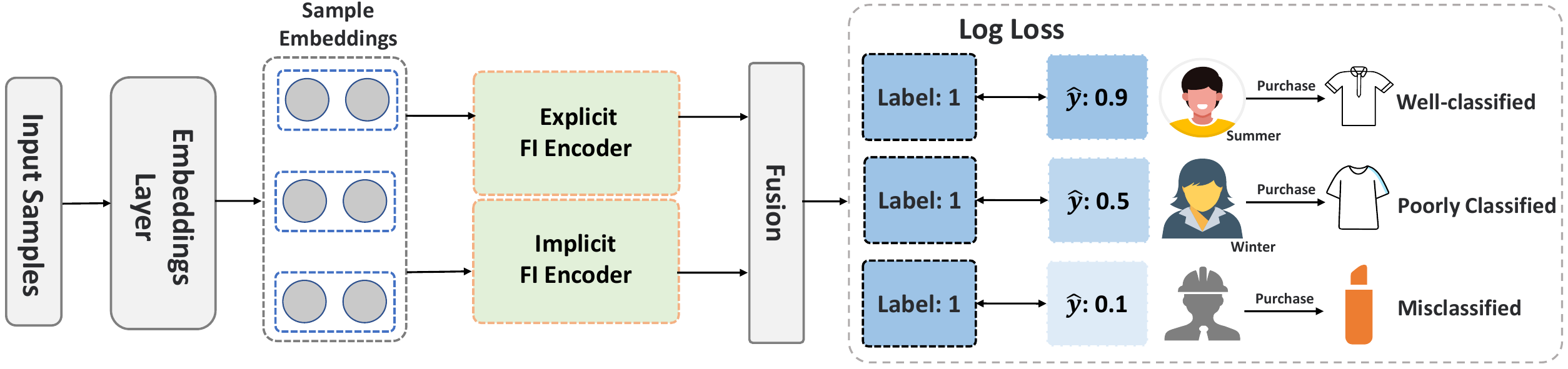}
    \end{minipage}
    \captionsetup{justification=raggedright}
    \caption{The workflow of current parallel-structured CTR models}
    \label{traditionallyCTR}
\end{figure}

The majority of CTR models \cite{dcn, EDCN, NFM, AFN, DCNv3} are dedicated to building diverse explicit Feature Interaction (FI) encoders that are product-based, which are then synergized with implicit FI encoders grounded in Multi-Layer Perceptron (MLP) to enhance performance.  The architecture of these models can be divided into stacked and parallel structures depending on their integration approach \cite{dcnv2, SimCEN}. A stacked structure \cite{pnn1,pnn2,DIN,DIEN} uses a sequential linking of explicit and implicit encoders, with the output of one encoder feeding into the next. In contrast, a parallel structure \cite{GDCN, CL4CTR, dcnv2}, as shown in Figure \ref{traditionallyCTR}, typically integrates encoders parallelly, allowing all encoders to learn simultaneously and integrate results at a fusion layer. 

\begin{figure*}[t]
\centering
\subfloat[Training on Frappe]
{\includegraphics[width=0.25\textwidth]{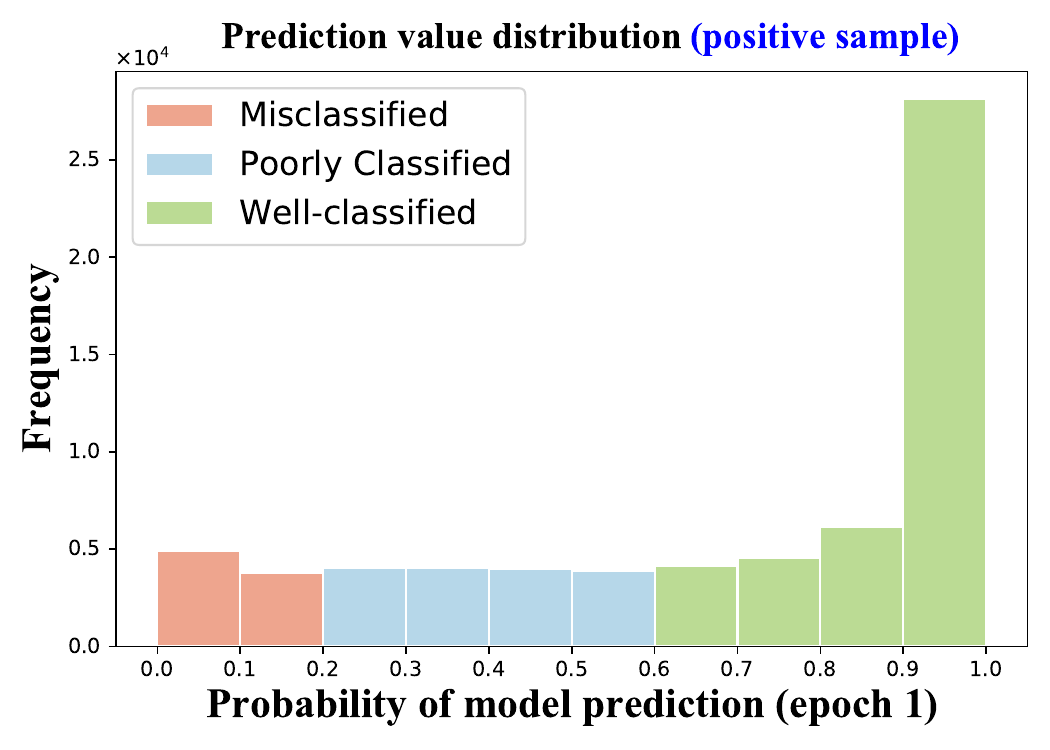}%
\label{a}}
\hfil
\subfloat[Training on Frappe]
{\includegraphics[width=0.25\textwidth]{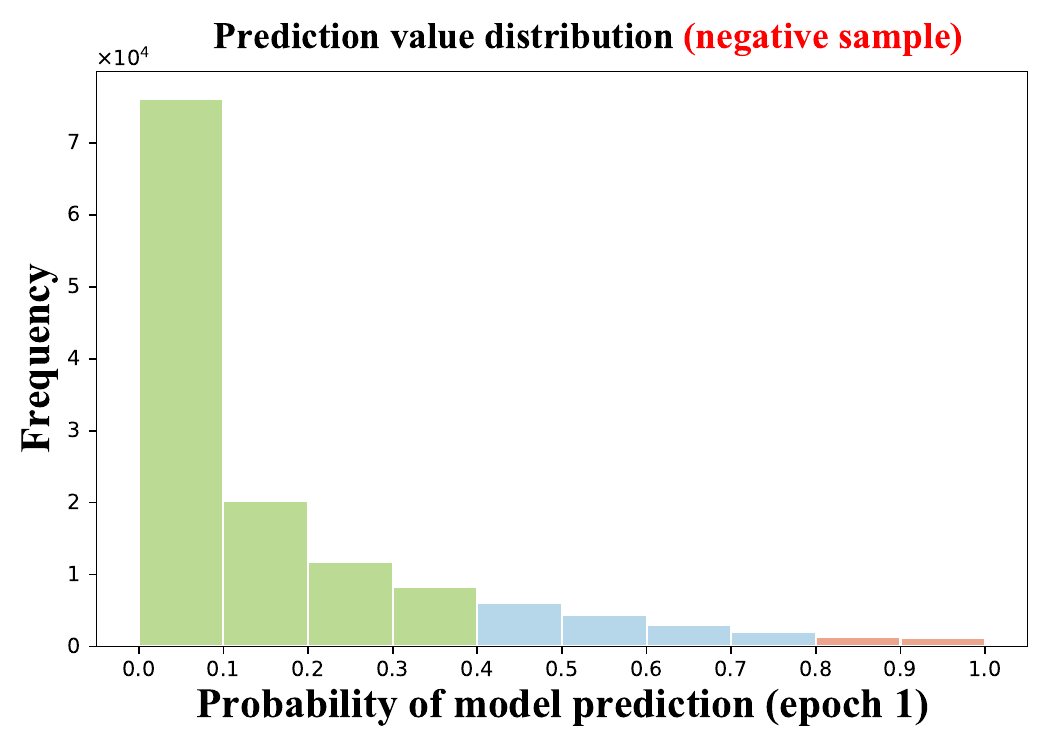}%
\label{b}}
\hfil
\subfloat[Training on ML-tag]
{\includegraphics[width=0.25\textwidth]{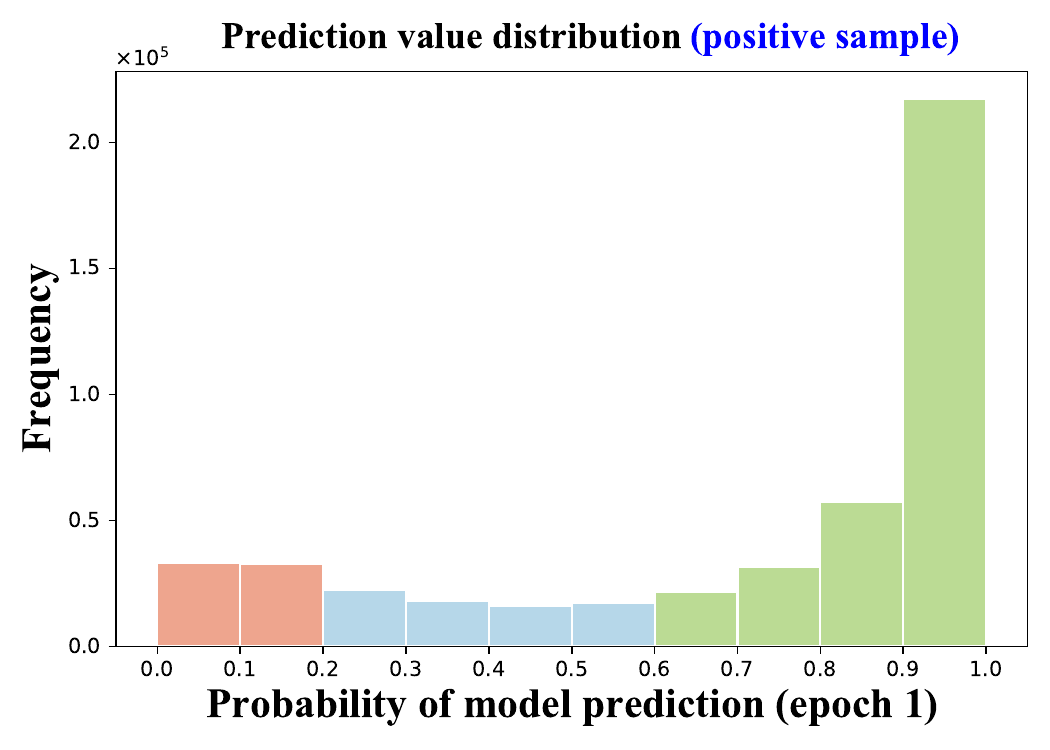}%
\label{c}}
\hfil
\subfloat[Training on ML-tag]
{\includegraphics[width=0.25\textwidth]{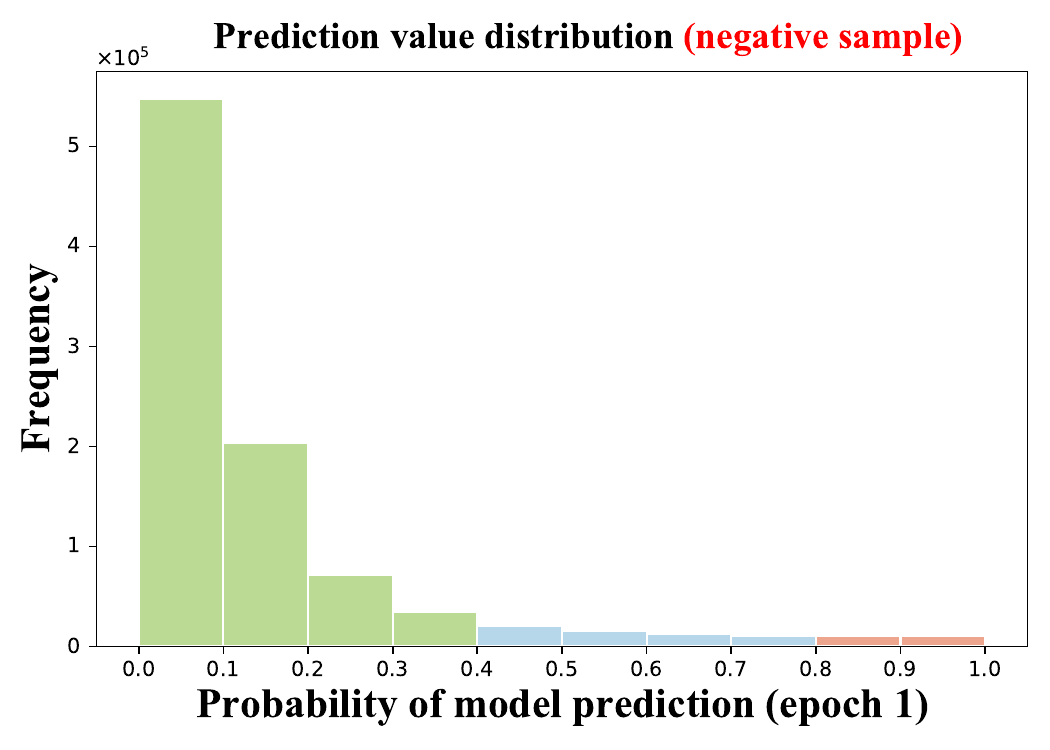}%
\label{d}}
\hfil
\subfloat[Validation on Frappe]{\includegraphics[width=0.25\textwidth]{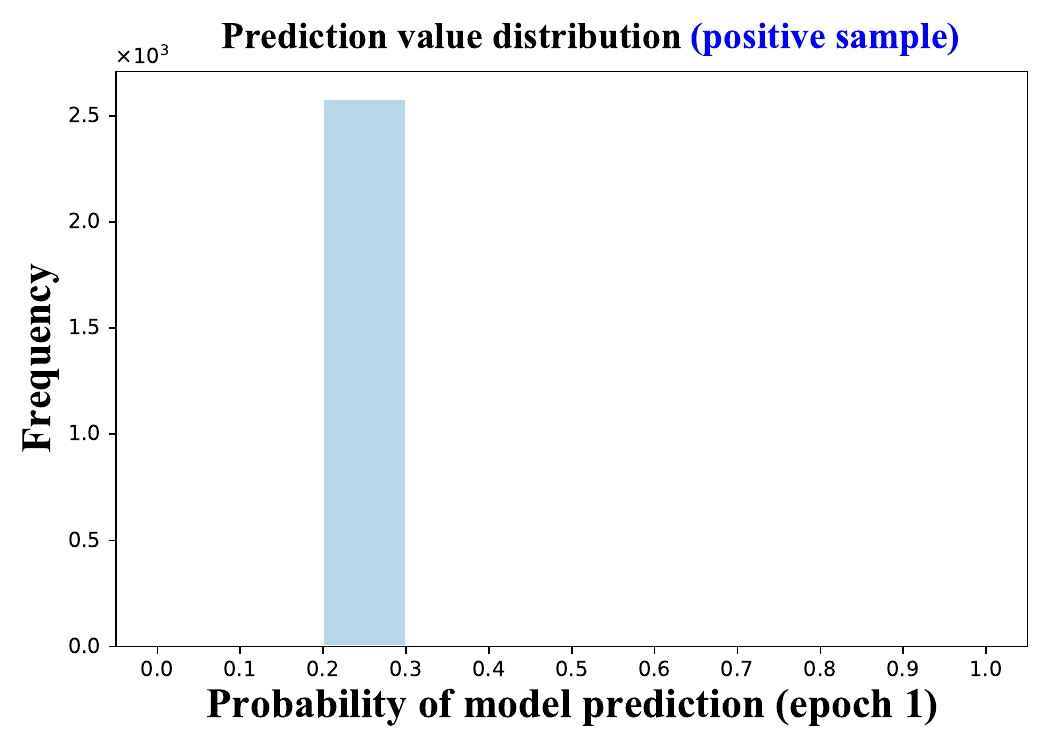}%
\label{e}}
\hfil
\subfloat[Validation on Frappe]{\includegraphics[width=0.25\textwidth]{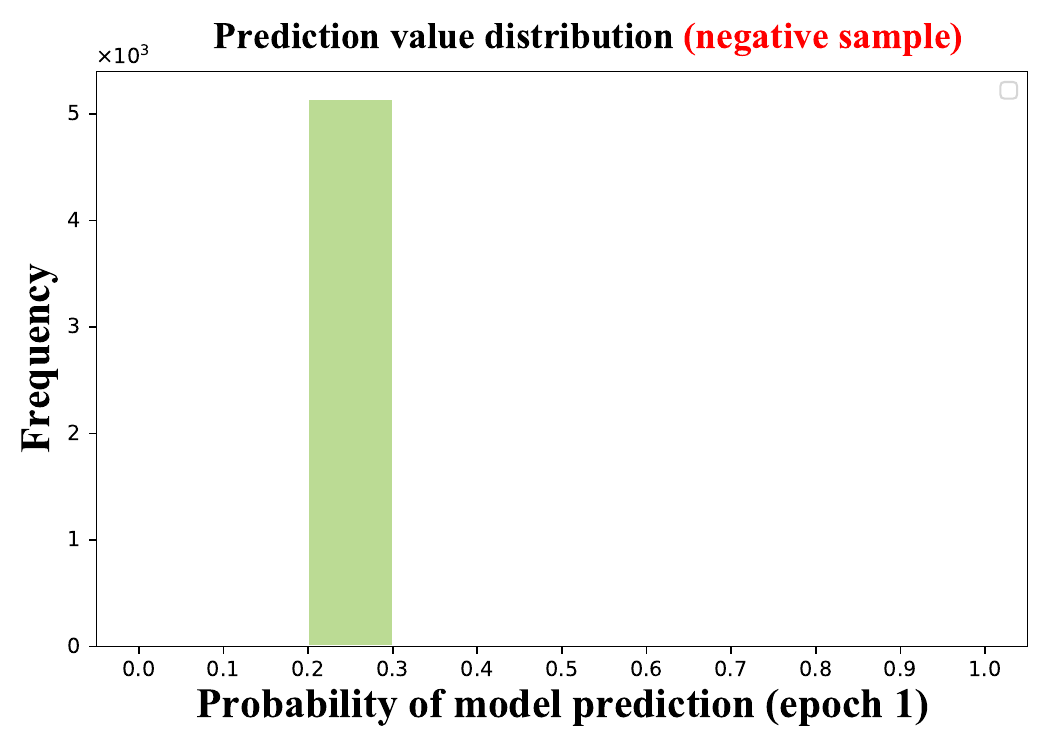}%
\label{f}}
\hfil
\subfloat[Validation on ML-tag]{\includegraphics[width=0.25\textwidth]{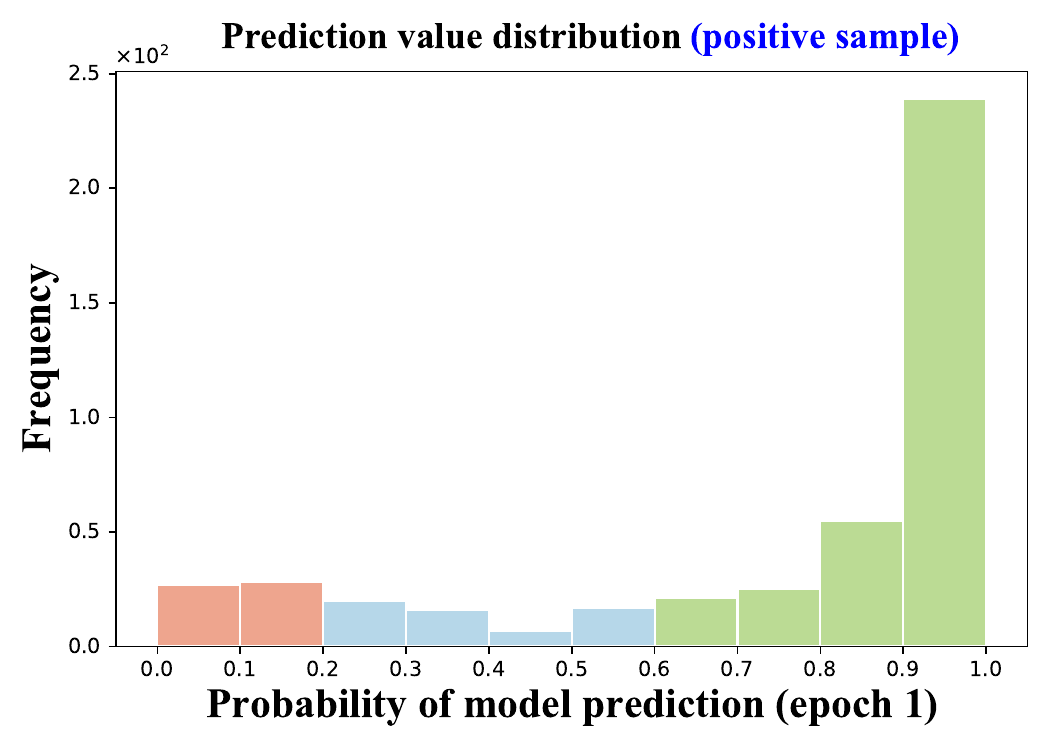}%
\label{g}}
\hfil
\subfloat[Validation on ML-tag]{\includegraphics[width=0.25\textwidth]{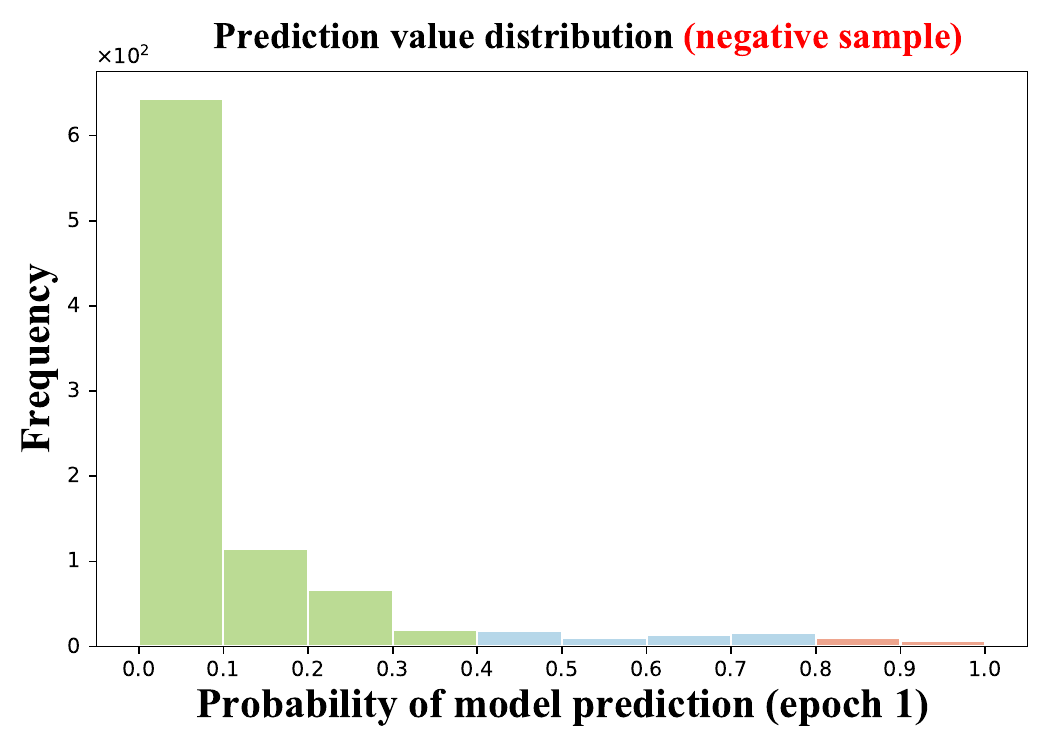}%
\label{h}}
\hfil
\subfloat[Validation on Frappe]{\includegraphics[width=0.25\textwidth]{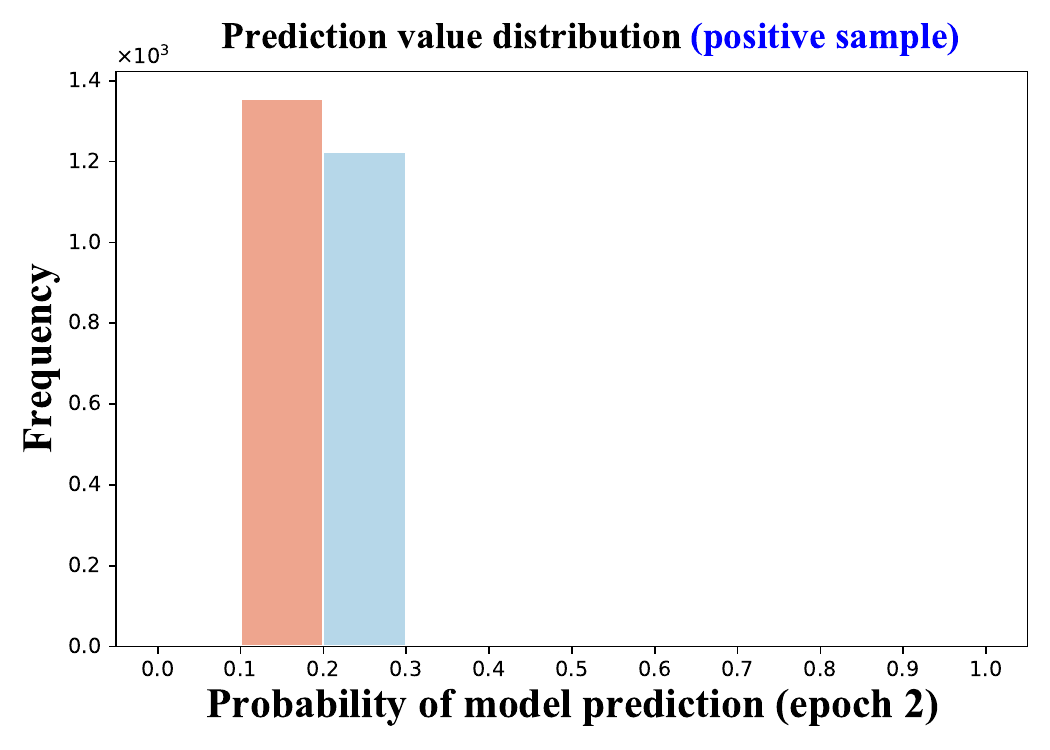}%
\label{i}}
\hfil
\subfloat[Validation on Frappe]{\includegraphics[width=0.25\textwidth]{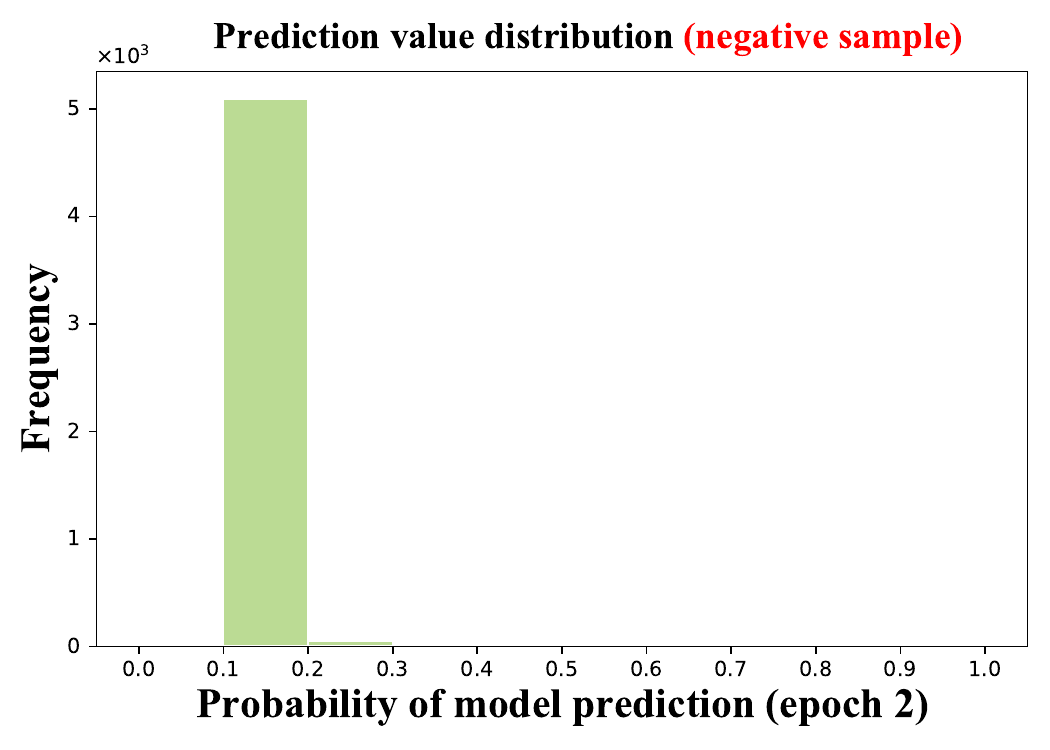}%
\label{j}}
\hfil
\subfloat[Validation on ML-tag]{\includegraphics[width=0.25\textwidth]{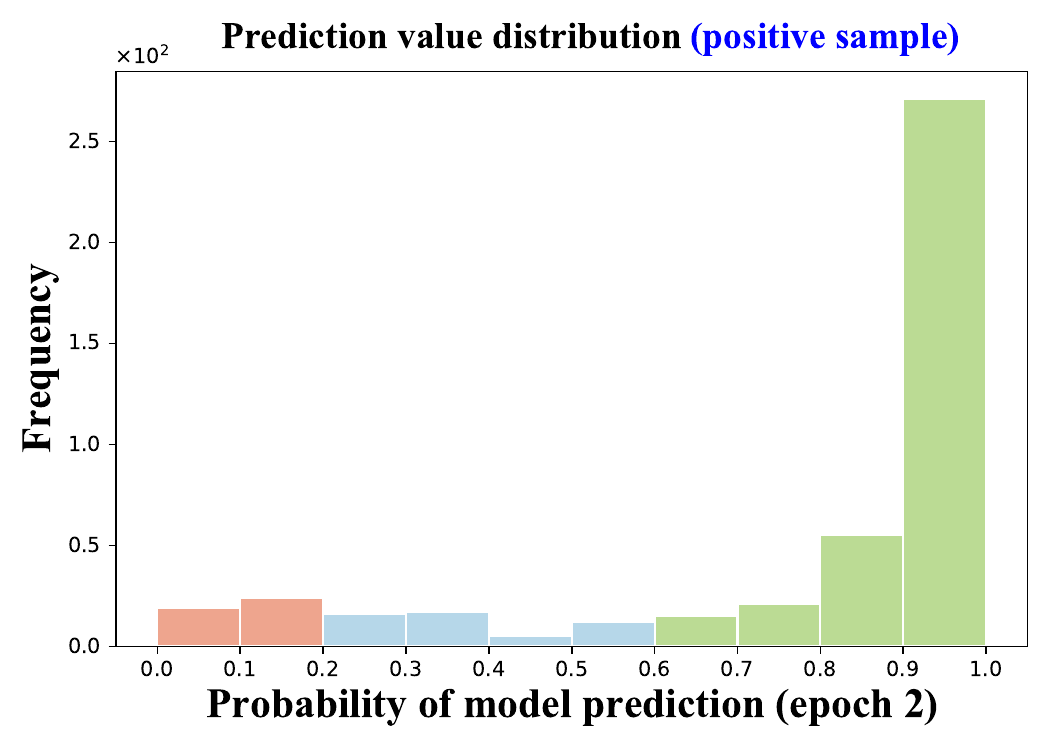}%
\label{k}}
\hfil
\subfloat[Validation on ML-tag]{\includegraphics[width=0.25\textwidth]{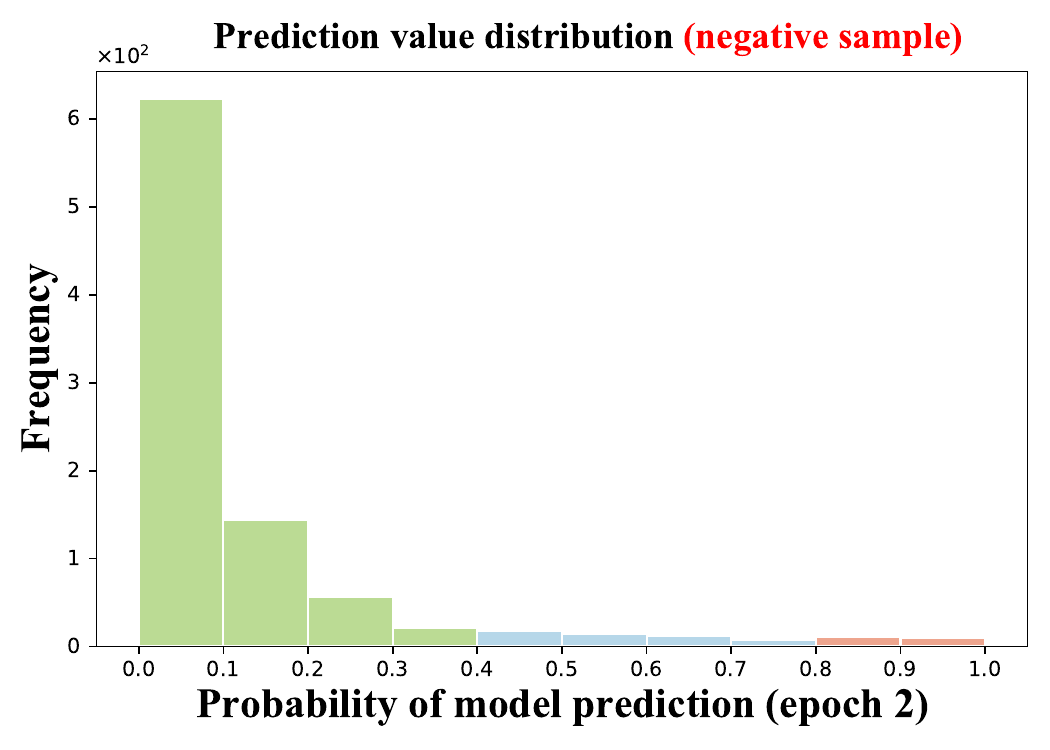}%
\label{l}}
\caption{Distribution of DNN's prediction results for positive and negative samples on training and validation datasets. Red indicates misclassified, blue indicates poorly classified, and green indicates well-classified.}
\label{Unbalanced}
\end{figure*}

Compared to stacked structures, the CTR model with parallel structure has received considerable research attention due to their decoupled ability and parallel computing-friendly properties \cite{EDCN, dcnv2, finalmlp}, leading to state-of-the-art (SOTA) performance \cite{GDCN,FINAL}. Despite the effectiveness of current CTR models based on this parallel structure, there are limitations to overcome:
\begin{itemize} 
\item \textbf{Insufficient sample differentiation ability.} As shown in Figure \ref{traditionallyCTR}, we classify the samples in terms of the gap between the model predictions and the labels as: \#\textit{Well-classified} (corresponding to easy sample, e.g., a male student purchasing a T-shirt during summer, aligning with typical behavior), \#\textit{Poorly Classified} (corresponding to challenging sample, e.g., a female student buys a T-shirt in winter, which seems unconventional but is explained by her spending a lot of time in warm indoor environments), and \#\textit{Misclassified} (corresponding to hard sample, e.g., a male engineer purchases lipstick, which is non-typical behavior for males). Most CTR models learn the above three classes of samples indiscriminately, without the ability to differentiate samples based on complexity or learning difficulty. Without this, models may overlook the informative hard samples essential for enhancing robustness and predictive accuracy \cite{hardsample}.

\item \textbf{Unbalanced sample distribution.} We visualize the distribution of the DNN's prediction results during the initial learning phase, as illustrated in Figure \ref{Unbalanced}. The CTR model fits most of the samples (i.e., more green part) quickly after the first epoch. However, multiple problems arise on the validation set: (1) Figure \ref{Unbalanced} (a $\sim$ d) reveals an obvious thing that there are significantly more easy samples than hard samples. This disproportionate influence of   easy samples can limit a model's effectiveness and its generalization capabilities \cite{focalloss}; (2) from Figure \ref{Unbalanced} (e) and (i), we can identify insufficient generalization on the Frappe dataset; (3) from Figure \ref{Unbalanced} (e $\sim$ l), the model does not fit  challenging and hard   samples well as the number of epochs increases.

\item \textbf{Undifferentiated learning process.} As we show in Figure \ref{traditionallyCTR}, the training process of most CTR models tends to use the single supervision signal for all encoders (i.e., Log Loss) \cite{dcnv2, GDCN, finalmlp}, ignoring the varying degree to which different samples contribute to the learning process. This may result in poor encoder training and failure to capture proper interaction information.
\end{itemize}

To address the above limitations, we introduce a novel CTR prediction framework, called the \textbf{Twin Focus Framework for CTR} (TF4CTR). Considering the excellent performance of parallel structures \cite{EDCN, openbenchmark}, we focus on applying our proposed framework within CTR models that utilize such structures, thereby helping various representative baseline models to differentiate sample complexity in a model-agnostic manner adaptively. This enables encoders to capture feature interaction information more effectively and target-specific. More specifically, this framework incorporates two plug-and-play modules alongside a target-specific loss function designed to enhance the model's prediction and generalization capabilities: (1) \textbf{Sample Selection Embedding Module (SSEM)} aims to select different encoders for samples of different complexity, thus preventing the encoder from being simultaneously biased towards capturing information in easy samples. SSEM is positioned at the base of the model, and adaptively selects the most appropriate encoder for the current input sample. (2) \textbf{Dynamic Fusion Module (DFM)} dynamically aggregates and evaluates the outputs of different encoders to achieve a more accurate prediction outcome. In this way, it can take advantage of multiple encoding strategies to dynamically synthesize information to accommodate the diversity of input data. (3) \textbf{Twin Focus (TF) Loss} provides target-specific auxiliary supervision signals to encoders of varying complexities. It aims to refine the learning process by providing differentiated training signals that emphasize the learning of complex encoders to harder samples or simple encoders to easy samples, thus reducing the effect of the imbalance in the number of hard and easy samples and improving the overall generalization ability of the model. The major contributions of this paper are summarized as follows:
\begin{itemize} 
\item We elucidate three inherent limitations of current parallel-structured CTR models and confirm their existence through experimental validation and theoretical analysis.
\item We propose a novel model-agnostic CTR framework, called TF4CTR, which improves the information capture ability of the encoder and the final prediction accuracy of the model through adaptive sample differentiation.
\item We introduce a lightweight Sample Selection Embedding Module, Dynamic Fusion Module, and Twin Focus Loss, all of which can be seamlessly integrated as plug-and-play components into various CTR baseline models to improve their generalization capabilities and performance.
\item We conduct comprehensive experiments across five real-world datasets, demonstrating the effectiveness and compatibility of the proposed TF4CTR framework.
\end{itemize}

\section{Preliminaries}
\subsection{CTR Prediction Task}
{
To facilitate the subsequent method description and theoretical analysis, we first summarize the main notations used throughout this paper in Table~\ref{tab:notation}.
}

\begin{table}[t]
{
\renewcommand\arraystretch{1.1}
\centering
\caption{Main Notations}
\label{tab:notation}
\scriptsize
\resizebox{\columnwidth}{!}{%
\begin{tabular}{c|l}
\hline
Symbol & Description \\
\hline
$X$ & Input sample (instance), a tuple of feature fields $\{x_p, x_a, x_c\}$ \\
$x_i$ & The $i$-th feature (categorical value) in a sample \\
$y \in \{0,1\}$ & Ground-truth click label ($1$ for click, $0$ otherwise) \\
$N$ & Mini-batch size \\
$f$ & Number of feature fields \\
$s_i$ & Vocabulary size of the $i$-th field \\
$d$ & Embedding dimension of each field \\
$D$ & Total embedding dimension, $D = \sum_{i=1}^{f} d$ \\
\hline
$E_i \in \mathbb{R}^{d \times s_i}$ & Embedding matrix for the $i$-th field \\
$\mathbf{e}_i \in \mathbb{R}^{d}$ & Embedding vector of feature $x_i$ \\
$\mathbf{h} \in \mathbb{R}^{D}$ & Concatenated embedding $\mathbf{h} = [\mathbf{e}_1,\dots,\mathbf{e}_f]$ \\
$\mathbf{E}_{es}$, $\mathbf{E}_{hs}$ & Embedding matrices in SSEM for the easy / hard branches \\
$\mathbf{h}_{es}$, $\mathbf{h}_{hs}$ & Input embeddings to the simple / complex FI encoders \\
$\hat{\mathbf{h}}_{es}$, $\hat{\mathbf{h}}_{hs}$ &
Latent representations output by the simple / complex FI encoders \\
\hline
$\texttt{MLP}_s$, $\texttt{MLP}_c$ & Simple FI encoder / complex FI encoder \\
$z_s$, $z_c$ & Logits output by the simple / complex FI encoders \\
$\hat{y}_s$, $\hat{y}_c \in (0,1)$ & Predicted CTR from the simple / complex FI encoders \\
$\hat{y} \in (0,1)$ & Final CTR prediction after fusion \\
\hline
$\mathrm{g}$ & Gating score in SSEM (scalar gate controlling the easy / hard branches) \\
$\mathbf{m}_1, \mathbf{m}_2$ & Outputs of expert networks in MMoE-style SSEM \\
$w_s, w_c$ & Learnable fusion weights in Weighted Sum Fusion (WSF) \\
$p_s, p_c$ & Selection probabilities in Voting Fusion (VF) \\
$\mathbf{W}_{cf}, b_{cf}$ & Parameters of the fusion layer in Concatenation Fusion (CF) \\
$g_1, g_2$ & Gating weights in Mixture-of-Experts Fusion (MoEF) \\
$m_1, m_2$ & Expert outputs in MoEF (logits before sigmoid) \\
\hline
$\mathcal{L}_{ctr}$ & Main CTR loss (binary cross-entropy) \\
$\mathcal{L}_{simple}$ & Auxiliary loss for the simple FI encoder \\
$\mathcal{L}_{complex}$ & Auxiliary loss for the complex FI encoder \\
$\mathcal{L}_{TF}$ & Twin Focus Loss, $\mathcal{L}_{TF} = \alpha \mathcal{L}_{simple} + (1-\alpha)\mathcal{L}_{complex}$ \\
$\mathcal{L}_{total}$ & Total training objective, $\mathcal{L}_{total} = \mathcal{L}_{ctr} + \mathcal{L}_{TF}$ \\
$\alpha$ & Trade-off coefficient between $\mathcal{L}_{simple}$ and $\mathcal{L}_{complex}$ \\
$c \in [0,1]$ & TF Loss hyperparameter controlling the emphasis on easy vs. hard samples \\
$m = 2-c$ & Complementary coefficient used in $\mathcal{L}_{complex}$ \\
$\gamma$ & Modulating factor of TF Loss (typically $\gamma \in \{1,2,3\}$) \\
\hline
\end{tabular}%
} 
}
\end{table}

CTR prediction is typically considered a binary classification task that utilizes user profiles \cite{openbenchmark, autoint}, item attributes, and context as features to predict the probability of a user clicking on an item. The composition of these three types of features is as follows:
\begin{itemize} 
\item  \emph{User profiles} ($p$): age, gender, occupation, etc.
\item \emph{Item attributes} ($a$): brand, price, category, etc.
\item \emph{Context} ($c$): timestamp, device, position, etc.
\end{itemize}
Further, we can define a CTR sample in the tuple data format: $X = \{x_p, x_a, x_c\}$. Variable $y \in \{0, 1\}$ is an true label for user click behavior:
\begin{equation}
    y= \begin{cases}1, & \text{user} \text { has clicked } \text{item}, \\ 0, & \text { otherwise. }\end{cases}
\end{equation}
It is a positive sample when $y=1$ and a negative sample when $y=0$. The final purpose of the CTR prediction model, which is to reduce the gap between the model's prediction and the true label, is formulated as follows:
\begin{equation}
\begin{aligned}
   &{\arg \min }\|y-\hat{y}\|,\ \ \hat{y}=\texttt{MODEL}(X).
\end{aligned}
\end{equation}
where \texttt{MODEL} denotes the CTR model.

\subsection{Three Categories of Samples}
In the real world, CTR prediction tasks handle millions of data points \cite{DIEN, AT4CTR} and encounter issues related to high sparsity and the cold start problem, with some datasets exhibiting sparsity levels exceeding 99.9\% \cite{autoint}. Consequently, the samples within these datasets inevitably present varied levels of prediction difficulty. Based on the discrepancies between the model's predictions and the true labels, we can categorize these samples into three categories:
\begin{itemize} 
\item \#\textit{Well-classified} $\leftrightarrow$  easy sample, 
\item \#\textit{Poorly Classified} $\leftrightarrow$ challenging sample, 
\item \#\textit{Misclassified} $\leftrightarrow$ hard sample.
\end{itemize}

Since the CTR model's predictions fall between 0 and 1, we can set thresholds to distinguish between three categories of predictions. For example, for positive samples ($y=1$), predictions in the range [0.6, 1.0] can be considered as \textit{Well-classified}, those in [0.2, 0.6) as \textit{Poorly Classified}, and those in [0.0, 0.2) as \textit{Misclassified}. 
It is important to note that these thresholds are only used for post-hoc analysis and visualization (e.g., in Fig.~\ref{Unbalanced}), in order to provide an intuitive illustration of ``easy / challenging / hard'' samples.
They are not used anywhere in the training pipeline of TF4CTR, and the actual modeling of sample difficulty during optimization is fully handled by the continuous form of the proposed Twin Focus Loss.

\begin{figure}[t]
    \begin{minipage}[t]{1\linewidth}
        \centering
        \includegraphics[width=\textwidth]{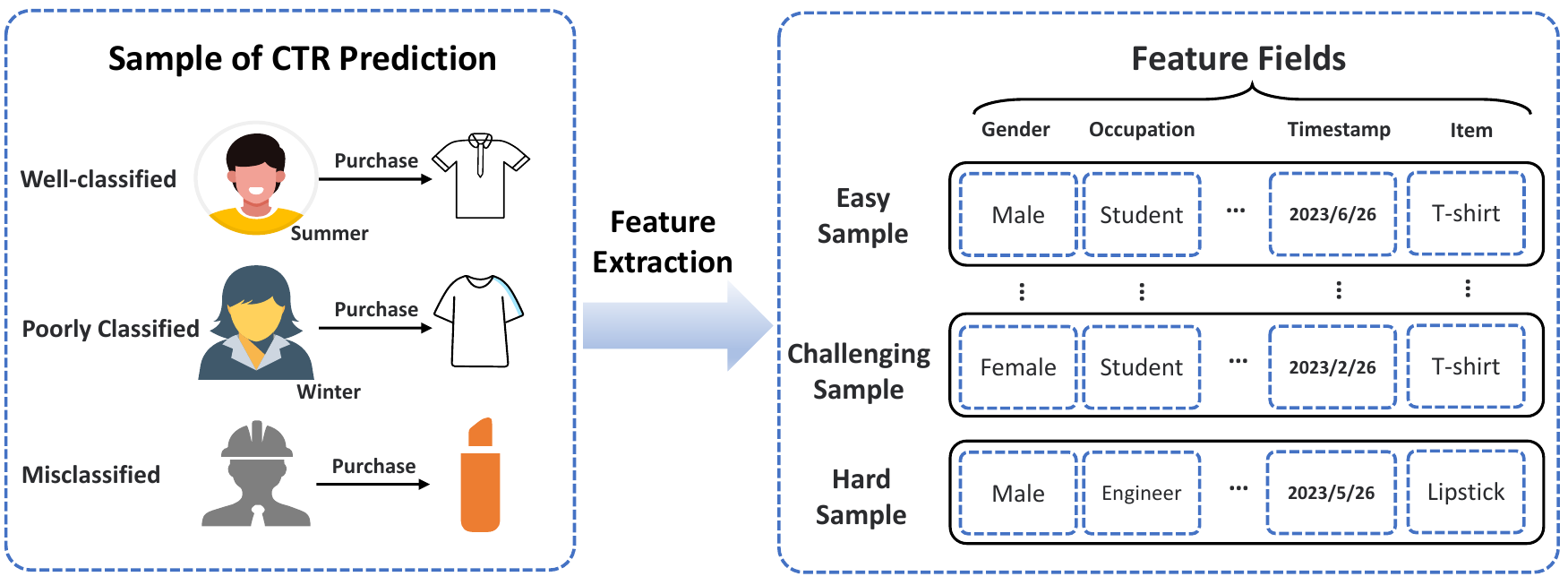}
    \end{minipage}
    \captionsetup{justification=raggedright}
    \caption{An example of a three-category sample.}
    \label{example}
\end{figure}

Typically, CTR prediction tasks randomly sample a mini-batch of $N$ user click behaviors. As depicted in Figure \ref{example}, each sample within the batch contains several feature fields $f$, with each field comprising its own attributes. Taking an easy sample as an example, \{\textit{Gender, Occupation, Timestamp, Item}\} $\in f$, we can extract the specific features of a click behavior: \{\textit{Male, Student, 6/26/2023, T-shirt}\}. From the interpretability perspective, it makes sense for a male student to show interest in purchasing a T-shirt in summer, thus it is categorized as an easy sample ($es$). In contrast, for the features set \{\textit{Male, Engineer, 5/26/2023, Lipstick}\}, this can be interpreted as a male engineer who suddenly decides to purchase lipstick in May, possibly as a gift. Because it is less typical for males to develop an interest in lipstick, this scenario would be classified as a hard sample ($hs$). For the challenging sample \{\textit{Female, Student, 2/26/2023, T-shirt}\}, it is observed that a female student purchases a T-shirt during winter. Although buying a T-shirt in winter may seem unconventional, the reason for this purchase could be that she spends significant time in heated indoor environments. Therefore, to more accurately predict user click behavior, it is essential to analyze user data patterns further.

\begin{figure*}[t]
    \begin{minipage}[t]{1\linewidth}
        \centering
        \includegraphics[width=\textwidth]{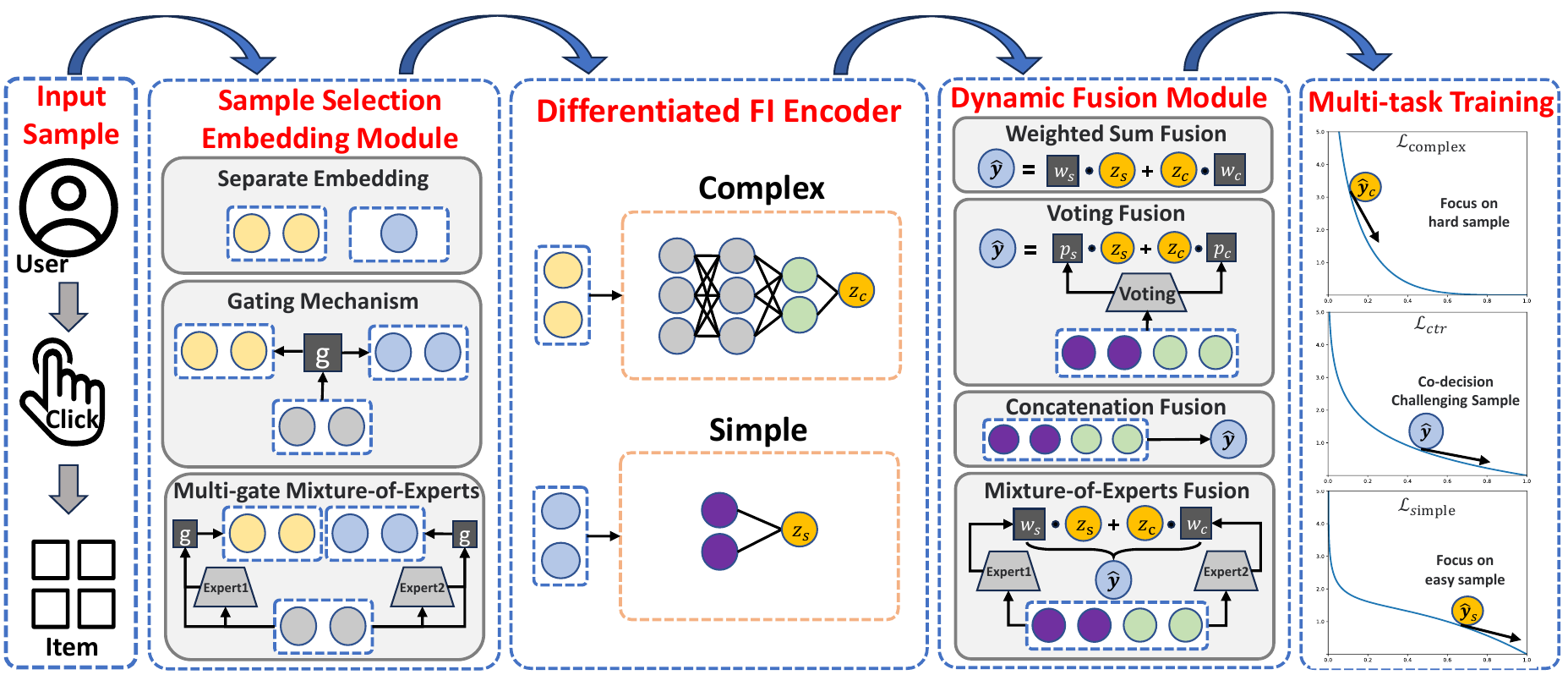}
    \end{minipage}
    \captionsetup{justification=raggedright}
    \caption{The overall framework of TF4CTR, comprising four distinct stages: (1) \textbf{Sample Selection}: A gating mechanism generates a score $g$ to dynamically categorize input samples into simple or complex encoder. (2) \textbf{Differentiated FI Encoder}: Two specialized experts process these subsets to generate latent representations ($\mathbf{z}_s, \mathbf{z}_c$) and initial predictions ($\hat{y}_s, \hat{y}_c$), guided by auxiliary losses $\mathcal{L}_{simple}$ and $\mathcal{L}_{complex}$. (3) \textbf{Dynamic Fusion Module (DFM)}: This module integrates the dual-stream information using one of four strategies: \textit{Weighted Sum Fusion} (linear combination), \textit{Voting Fusion} (probabilistic selection), \textit{Concatenation Fusion} (merging latent vectors $[\mathbf{z}_s; \mathbf{z}_c]$), or \textit{Mixture-of-Experts Fusion} (dynamic gating). (4) \textbf{Multi-task Training}: The final prediction is optimized via a joint loss function combining the main CTR loss $\mathcal{L}_{ctr}$ and auxiliary losses $\mathcal{L}_{TF}$.}
    \label{TF4CTR_frame}
\end{figure*}

\section{Proposed Method}
\subsection{Twin Focus Framework}
The TF4CTR framework introduces an adaptive strategy for differentiating between sample data. This allows for targeted learning of network parameters to predict CTR. Meanwhile, the core idea of TF4CTR aligns with the previous WideDeep \cite{widedeep}. The simple FI Encoder can be viewed as the Wide component for memorizing simple feature interactions, while the complex FI Encoder corresponds to the Deep component for enhancing the model's generalization. The introduction of TF Loss further strengthens this from the perspective of the supervision signal. As depicted in Figure \ref{TF4CTR_frame}, the architecture of TF4CTR is structured four main components:

\begin{itemize} 
\item \textbf{Sample Selection Embedding Module (SSEM)}: This component adaptively differentiates the post-embedding samples, directing them to the most appropriate encoder. For instance, easy samples are channeled to a simple encoder, hard samples are routed to a complex encoder, and challenging samples are determined collectively by multiple encoders.

\item \textbf{Differentiated Feature Interaction Encoder (FI Encoder)}: This encoder performs interactions on data samples to varying extents. Our proposed framework allows a variety of FI Encoders to achieve more precise predictions and enhance the model's overall generalization capability. To keep it simple, we adopt the simple and widely-used MLP as our base FI Encoder. The depth of the MLP is used to differentiate between the complex and simple FI Encoders, thereby generating two distinct predictions for the same sample.
\item \textbf{Twin Focus Loss (TF Loss)}: This loss function provides auxiliary supervision signals to the complex and simple FI Encoders. This encourages the encoders to learn target-specific feature interactions of complex (or simple) samples. 
\item \textbf{Dynamic Fusion Module (DFM)}: This module dynamically integrates the prediction results from the distinct FI Encoders to reduce classification errors. 
\end{itemize}

\subsection{Embedding Layer}
In most deep learning-based CTR models, the embedding layer is an essential component \cite{GDCN, finalmlp, autoint}. Simply put, for each input sample $x_i \in X$ ($x_i$ denotes a particular feature of the input sample $X$), it can be transformed into a high-dimensional sparse vector using one-hot encoding \cite{openbenchmark, fignn, CETN}, which is then converted into a low-dimensional dense embedding through the embedding matrix: 
$$\mathbf{e}_i = E_i x_i,$$
where $E_i \in \mathbb{R}^{d \times s_i}$ and $s_i$ separately indicate the embedding matrix and the vocabulary size for the $i^{th}$ field, and $d$ represents the embedding dimension. After that, we concatenate the individual features to get the input to the FI Encoder:
$$\mathbf{h}=\left[\mathbf{e}_1, \mathbf{e}_2, \cdots, \mathbf{e}_f\right] \in \mathbb{R}^{D},$$
where $f$ denotes the number of feature fields, and $D=\sum_{i=1}^f d$. However, it is often not a good choice to directly input $\mathbf{h}$ after the embedding layer into the FI Encoder, and the validity of this idea has been confirmed in many previous works \cite{GemNN, FINAL, pnn1, pnn2}.

\subsection{Sample Selection Embedding Module}
\label{SSEM}
In this work, we start from the predictive difficulty of the samples themselves and use the SSEM to make an adaptive selection. This module is designed to evaluate the complexity of the samples and guide them to different encoders in the model architecture. For example, easy samples like the previously mentioned male student interested in a T-shirt may be processed through a simpler encoder that captures the most straightforward features, while hard samples like the male engineer interested in lipstick may require a more sophisticated encoder that can learn to understand deeper data patterns and feature interactions. This adaptive approach aims to optimize the model's performance by ensuring that each sample is processed in an efficient and effective manner. By recognizing and treating samples differently based on their complexity, the SSEM contributes to an overall more accurate and robust CTR prediction model.

Specifically, we have defined several methods to implement the SSEM, as follows:
\begin{itemize} 
\item \textit{Separate Embedding Representations (SER):} Instead of sharing feature embeddings between different encoders \cite{deepfm}, we use two distinct embedding matrices $\mathbf{E}_{es}$ and $\mathbf{E}_{hs}$ to embed the input sample $X$, which greatly alleviates the problem of excessive sharing \cite{EDCN}. Additionally, by changing the embedding dimensions, samples can obtain representations that are better suited to their complexity. This is formalized as:
\begin{equation}
\begin{aligned}
    \mathbf{h}_{es} = \mathbf{E}_{es} X,\ \ \mathbf{h}_{hs} = \mathbf{E}_{hs} X,
\end{aligned}
\end{equation}
where $\mathbf{E}_{es} \in \mathbb{R}^{\frac{d}{2} \times s_i \times f}$ and $\mathbf{E}_{hs} \in \mathbb{R}^{d \times s_i \times f}$, $\mathbf{h}_{es}$ and $\mathbf{h}_{hs}$ denote easy sample embedding and hard sample embedding respectively.

\item \textit{Gating Mechanism (GM):} A simple yet effective gating network can serve as a judge to direct samples of varying difficulties to appropriate encoders. To reduce the number of model parameters, a single gating unit is utilized to control the input embeddings for FI Encoders:
\begin{equation}
\begin{aligned}
    \mathrm{g} &= \sigma(\texttt{Gate}(\mathbf{h})),\\
    \mathbf{h}_{es} &= \mathrm{g} \cdot \mathbf{h}, \ \ \mathbf{h}_{hs} = (1 - \mathrm{g}) \cdot  \mathbf{h},
\end{aligned}
\end{equation}
where $\mathrm{g} \in \mathbb{R}^1$ is the gate value, $\sigma$ is the sigmoid function, $\cdot$ denotes the dot product, and \texttt{Gate} denotes a gating unit. For convenience, we implement this using a simple MLP, and other more complex gating networks could be used here.
\item \textit{Multi-gate Mixture-of-Experts Structures (MMoE):} To address the potential issue of insufficient sample differentiation capability with a gating unit alone, we further incorporate the concept of MMOE \cite{mmoe}. This approach allows multiple expert networks to evaluate the same sample and selects the most suitable FI Encoder for the sample using the gating mechanism. Considering the complexity, we only use two experts paired with their corresponding gating units:
\begin{equation}
\begin{aligned}
    \mathrm{g}_{11}, \mathrm{g}_{12} &= \texttt{SoftMax}(\texttt{Gate}_1(\mathbf{h})),\\
    \mathrm{g}_{21}, \mathrm{g}_{22} &= \texttt{SoftMax}(\texttt{Gate}_2(\mathbf{h})), \\
    \mathrm{\textbf{m}_1} &= \texttt{EX}_1(\mathbf{h}),\ \ \mathrm{\textbf{m}_2} = \texttt{EX}_2(\mathbf{h}),\\
    \mathbf{h}_{es} &= \mathrm{{g}_{11}} \cdot  \mathrm{\textbf{m}_1} + \mathrm{{g}_{12}} \cdot \mathrm{\textbf{m}_2},\\
    \mathbf{h}_{hs} &= \mathrm{{g}_{21}} \cdot \mathrm{\textbf{m}_1} + \mathrm{{g}_{22}} \cdot \mathrm{\textbf{m}_2},
\end{aligned}
\end{equation}
where \texttt{EX} represents an expert network, $\mathrm{\textbf{m}} \in \mathbb{R}^{D}$ represents the output of the expert network. 
\end{itemize}


\subsection{Differentiated FI Encoder}
\label{FI Encoder}
 Improving the FI Encoder has always been a focal point of research in the CTR prediction \cite{AFN, dcnv2, xdeepfm, EulerNet, EDCN}. Generally, FI Encoders can be categorized into two types based on their encoding methods: (1) explicit FI Encoders \cite{autoint, GDCN} enhance the order of feature interactions according to predefined rules, thereby obtaining interpretable low-order feature interactions; (2) Implicit FI Encoders \cite{finalmlp, CL4CTR} capture the interactions between features implicitly through the deep neural network. Most studies \cite{deepfm,dcnv2,GDCN} suggest that combining explicit and implicit FI Encoders can further improve model performance. As a result, researchers are dedicated to designing various more complex explicit FI Encoders to enhance the overall predictive ability and interpretability of the model.

In this work, however, we categorize FI Encoders simply as either simple or complex encoders, allowing samples of varying prediction difficulty levels to have more suitable differentiated encoding pathways. To keep it simple, we use MLP with different numbers of layers to validate this idea.
\begin{equation}
\begin{aligned}
    \hat{\mathbf{h}}_{es} &= \texttt{MLP}_{s}(\mathbf{h}_{es}),\ \ \hat{\mathbf{h}}_{hs} = \texttt{MLP}_{c}(\mathbf{h}_{hs}),\\
z_s&=\mathbf{W}_{s}\hat{\mathbf{h}}_{es},\ \ z_c=\mathbf{W}_{c}\hat{\mathbf{h}}_{hs},\\
    \hat{y}_s&=\sigma(z_s),\ \ \hat{y}_c=\sigma(z_c),
\end{aligned}
\end{equation}
 where $\hat{\mathbf{h}}_{es}$ ($\hat{\mathbf{h}}_{hs}$) denotes the final representation of the encoder, the number of layers in $\texttt{MLP}_{s}$ is smaller than the number of layers in $\texttt{MLP}_{c}$, $\hat{y}_e$ and $\hat{y}_c$ denote the prediction results of the simple encoder ($\texttt{MLP}_{s}$) and complex encoder ($\texttt{MLP}_{c}$) respectively, $\mathbf{W}_s$ and $\mathbf{W}_c$ are the learned weights, and $z$ is the logit value. Other more complex or simpler FI Encoders could similarly be employed.

\subsection{Twin Focus Loss Function}
\begin{figure}[t]
    \begin{minipage}[t]{1\linewidth}
        \centering
        \includegraphics[width=\textwidth]{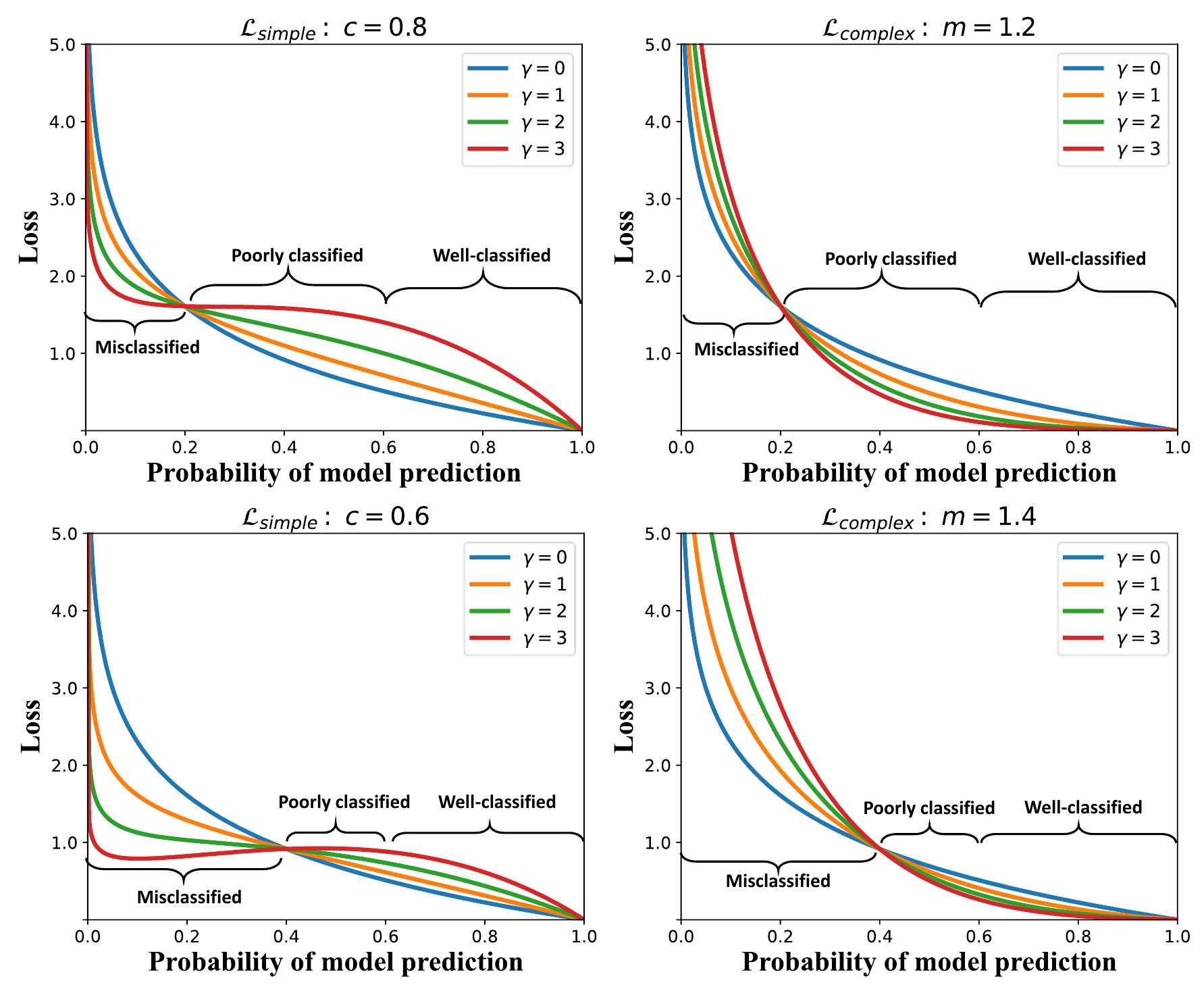}
    \end{minipage}%
    \captionsetup{justification=raggedright}
    \caption{ Visualization of the proposed $\mathcal{L}_{TF}$ under various hyperparameter settings. The curves illustrate the gradient magnitude across different prediction confidence levels, categorized into three regions: \textbf{Misclassified}, \textbf{Poorly classified}, and \textbf{Well-classified}. $\mathcal{L}_{simple}$ suppresses gradients within the \textbf{Misclassified} region while amplifying them in the remaining regions, thereby encouraging the encoder to prioritize easy and challenging samples. Conversely, $\mathcal{L}_{complex}$ suppresses gradients in the \textbf{Poorly classified} and \textbf{Well-classified} regions, directing the encoder to focus on hard samples.}
    \label{TFLOSS}
\end{figure}

Relying only on a single and undifferentiated loss function for specialized training of differentiated FI Encoders is undoubtedly inefficient, and introducing an auxiliary loss function is an effective strategy \cite{DIEN}. Therefore, we further add an auxiliary loss to the FI Encoders, which is designed to facilitate the learning of parameters by both simple FI Encoders and complex FI Encoders that are more favorable to their respective target sample classes. Doing so improves the model's ability to specialize on samples with different levels of difficulty. More specifically, this loss function encourages the simple FI Encoder to better fit easy samples, while encouraging the complex FI Encoder to better fit hard samples, and for the challenged samples to be jointly determined by the two FI Encoders. The general form of TF Loss is shown below:
\begin{equation}
\label{eq_tfloss}
\begin{aligned}
    &\mathcal{L}_{T F}=\alpha\mathcal{L}_{\text{simple}} + (1 - \alpha)\mathcal{L}_{\text{complex}}, \\
    &\mathcal{L}_{\text{simple}}=-\frac{1}{N} \sum_{i=1}^N (c+\hat{y}_{s,i})^\gamma \log \left(\hat{y}_{s,i}\right),\ \ c \in [0, 1] \\
    &\mathcal{L}_{\text{complex}}=-\frac{1}{N} \sum_{i=1}^N (m-\hat{y}_{c,i})^\gamma \log \left(\hat{y}_{c,i}\right),\ \ m=2-c \\
\end{aligned}
\end{equation}
where the formulation above corresponds to the loss for positive samples.
Here, $\alpha$ controls the balance between the losses attributed to the simple and complex FI Encoders, $c \in [0, 1]$ and $m \in [1, 2]$ are a pair of coefficients. The term $\gamma$ acts as a modulating factor and is typically selected from set $\{1, 2, 3\}$. These hyperparameters allow for the fine-tuning of the TF Loss, ensuring that it is suitably calibrated to the specific requirements of the sample difficulty being addressed. $\mathcal{L}_{T F}$ is visualized for several hyperparameters in Figure \ref{TFLOSS}. Briefly, the hyperparameter $c$ (and its corresponding $m$) can be used as a threshold to divide the range of applicability of various prediction results, thereby adjusting the corresponding loss function crosspoints. $\gamma$ then further adjusts the size of the loss using the $c$ dividing result as a reference.

Intuitively, as illustrated in Figure \ref{TFLOSS}, it is evident that a decrease in $c$ results in expanded applicability for \#\textit{Misclassified} samples, while simultaneously diminishing the scope for \#\textit{Poorly classified} and \#\textit{Well-classified} samples. Concurrently, as $\gamma$ increases, the loss corresponding to the \#\textit{Misclassified} range decreases for $\mathcal{L}_{simple}$, and the losses for \#\textit{Poorly classified} and \#\textit{Well-classified} samples increase. This means that $\mathcal{L}_{simple}$ guided FI Encoder will be more inclined to learn \#\textit{Poorly classified} and \#\textit{Well-classified} samples. Conversely, the opposite trend is observed for $\mathcal{L}_{complex}$. By manipulating the loss function in this manner, we enable it to exhibit differential sensitivity to various types of samples, thereby guiding the two distinct FI Encoders to refine their optimization process for the sample types they are inherently more proficient at handling. This approach helps to enhance the performance of the model and its ability to generalize to sophisticated samples.

{
From the perspective of gradient behavior and learning dynamics, our TF Loss is fundamentally different from representative difficulty-aware objectives such as Focal Loss and R-CE.
Focal Loss introduces a single-head difficulty re-weighting by multiplying the standard cross-entropy gradient with a modulating factor $(1-p_t)^{\gamma_f}\in(0,1]$, where $p_t$ denotes the predicted probability of the ground-truth class.
This mechanism suppresses gradients from well-classified (easy) samples and relatively emphasizes harder ones.
However, the re-weighting is shared across model components and does not provide encoder-specific supervision; therefore, in parallel-structured CTR models, different encoders are still trained under essentially consistent objectives, only with a uniform difficulty-based scaling.
R-CE can be viewed as a robust training strategy that further down-weights (or effectively truncates) gradients from extreme-hard samples that may be treated as outliers/noise, which can stabilize optimization under noisy labels.
In CTR prediction, however, hard samples are often informative long-tail behaviors rather than annotation noise; hence overly discarding these samples may be sub-optimal for generalization.
In contrast, TF Loss performs complementary, encoder-specific gradient shaping to encourage specialization.
For a positive sample, when $\sigma(z)$ is small (hard sample), the simple-encoder gradient is down-scaled by $(c+\sigma(z_s))^{\gamma-1}<1$ while the complex-encoder gradient is up-scaled by $(m-\sigma(z_c))^{\gamma-1}>1$ for $\gamma>1$ (and vice versa for easy samples).
This complementary rescaling makes the simple encoder focus more on easy/challenging cases, while the complex encoder focuses more on hard cases, thereby avoiding the undifferentiated learning dynamics caused by using a single shared loss.
}

\subsection{Dynamic Fusion Module}
\label{DFM}
To effectively synthesize the predictions from the differentiated encoders, we propose a Dynamic Fusion Module (DFM). The DFM aims to integrate the outputs from the simple FI Encoder and the complex FI Encoder to derive the final prediction. We investigate four distinct fusion mechanisms to identify the optimal integration strategy:

\begin{itemize} 
 \item {\textit{Weighted Sum Fusion (WSF)}: This strategy assumes that the relative importance of each encoder is global. We introduce two learnable scalar parameters to weight the contribution of the logits from each encoder. The final prediction is computed as:
\begin{equation}
\begin{aligned}
\hat{y}=\sigma(w_s \cdot z_s + w_c \cdot z_c),
\end{aligned}
\end{equation}
where $w_s, w_c \in \mathbb{R}^{1}$ are learnable weights initialized to 0.5. This allows the model to learn a global preference for the encoder that offers better generalizability.}

\item { \textit{Voting Fusion (VF)}: Voting Fusion treats the encoders as an ensemble. To ensure the framework remains end-to-end differentiable, we employ the Gumbel-Softmax technique \cite{gumbel-softmax} instead of non-differentiable hard voting. We generate a soft selection vector via:
\begin{equation}
\begin{aligned}
    \pi_s &= \mathbf{W}_{es}\hat{\mathbf{h}}_{es},\ \ 
    \pi_c = \mathbf{W}_{hs}\hat{\mathbf{h}}_{hs},\\
    p_k &=\frac{\exp \left(\left(\log \left(\pi_k\right)+g_k\right) / \tau\right)}{\sum_{j \in \{s, c\}} \exp \left(\left(\log \left(\pi_j\right)+g_j\right) / \tau\right)}, \\
    \hat{y}&=\sigma(p_s \cdot z_s + p_c \cdot z_c),
\end{aligned}
\end{equation}
where $\mathbf{W}_{es}$ and $\mathbf{W}_{hs}$ are projection matrices, $p_k$ represents the selection probability for the $k$-th encoder ($k \in \{s, c\}$), $g_k$ are i.i.d samples drawn from Gumbel$(0, 1)$, and $\tau$ is the temperature coefficient controlling the smoothness of the distribution. This approximates categorical sampling, allowing gradients to propagate during training.}

\item \textit{Concatenation Fusion (CF)}: Unlike the previous methods that operate on prediction logits, Concatenation Fusion integrates the latent representations. We concatenate the feature vectors extracted by the differentiated encoders and project them through a linear layer:
\begin{equation}
\begin{aligned}
\hat{y}=\sigma\left(\mathbf{W}_{cf}\left[\hat{\mathbf{h}}_{es}\ ||\  \hat{\mathbf{h}}_{hs}\right] + b_{cf}\right),
\end{aligned}
\end{equation}
where $\mathbf{W}_{cf}$ is the learnable weight matrix, $b_{cf}$ is the bias, and $||$ denotes the concatenation operation. This method enables the model to capture non-linear interactions between the features extracted by different views before making the final decision.

\item \textit{Mixture-of-Experts Fusion (MoEF)}: To achieve instance-level dynamic integration, we adopt a Mixture-of-Experts mechanism. Here, the gating network determines the contribution of expert networks based on the concatenated representations:
\begin{equation}
\begin{aligned}
    \mathrm{g}_1, \mathrm{g}_2 &= \texttt{SoftMax}(\texttt{gate}\left[\hat{\mathbf{h}}_{es}\ ||\  \hat{\mathbf{h}}_{hs}\right]), \\
    \mathrm{{m}_1} &= \texttt{ex}_1(\left[\hat{\mathbf{h}}_{es}\ ||\  \hat{\mathbf{h}}_{hs}\right]),\\
    \mathrm{{m}_2} &= \texttt{ex}_2(\left[\hat{\mathbf{h}}_{es}\ ||\  \hat{\mathbf{h}}_{hs}\right]),\\
    \hat{y} &= \sigma(\mathrm{{g}_1} \cdot \mathrm{{m}_1} + \mathrm{{g}_2} \cdot \mathrm{{m}_2}),
\end{aligned}
\end{equation}
where $\texttt{gate}$ and $\texttt{ex}$ denote the micro gating unit and expert networks respectively, and $\mathrm{m} \in \mathbb{R}^1$ is the expert output. This mechanism allows the DFM to adaptively rely on specific experts for different input patterns.
\end{itemize}

\subsection{Multi-task Training}
To integrate the TF4CTR framework into the CTR prediction scenario, we employ a multi-task training strategy to co-optimize the TF Loss and the original CTR prediction loss in an end-to-end manner. Thus, the final objective function can be expressed as:
\begin{equation}
\label{eq_total_loss}
\begin{aligned}
\mathcal{L}_{total}&=\mathcal{L}_{ctr} + \mathcal{L}_{TF},
\end{aligned}
\end{equation}
where $\mathcal{L}_{\text{ctr}} = -\frac{1}{N}\sum_{i=1}^N \log(\hat{y}_{t,i})$, $\hat{y}_{t,i} = 1-\hat{y}_i$ if $y_i = 0$, and $\hat{y}_{t,i} = \hat{y}_i$ if $y_i = 1$. 
To clearly delineate the overall workflow of the TF4CTR, we provide a detailed description of its complete training process, as shown in Algorithm \ref{training_process}.

\begin{algorithm}[h]
{
\footnotesize
\caption{The training process of TF4CTR\label{training_process}}
  \SetAlgoLined 
  \KwIn{input samples $X \in N$;}
  \KwOut{model parameters $\Theta$; }
  Initialize parameters $\Theta$, optimizer, and learning rate scheduler\;
  \While{TF4CTR has not reached the early stopping patience threshold}{
    \For{each mini-batch $X \subset N$}{
        \tcp{(1) Forward propagation}
        $\mathbf{h}_{es},\ \mathbf{h}_{hs}$ $\leftarrow$ SSEM($X$) according to Section~\ref{SSEM}\;
        Get $\hat{y}_{s},\ \hat{y}_{c}, z_s, z_c$ from Differentiated FI Encoders according to Section~\ref{FI Encoder}\;
        $\hat{y}$ $\leftarrow$ DFM($z_s, z_c$) according to Section~\ref{DFM}\;
        Calculate Twin Focus Loss $\mathcal{L}_{TF}$ according to Eq.~(\ref{eq_tfloss})\;
        Calculate CTR Loss $\mathcal{L}_{ctr}$\;
        Get total loss $\mathcal{L}_{total}$ according to Eq.~(\ref{eq_total_loss})\;
        
        \tcp{(2) Backward propagation and optimizer step}
        Zero the gradients of all parameters\;
        Backpropagate $\mathcal{L}_{total}$ to obtain $\nabla_{\Theta} \mathcal{L}_{total}$\;
        Apply gradient clipping on $\nabla_{\Theta} \mathcal{L}_{total}$\;
        Update parameters $\Theta$ using the optimizer step\;
        Update the learning rate using the scheduler step\;
    }
  }
  \Return model parameters $\Theta$;
}
\end{algorithm}

First, we initialize the model parameters $\Theta$ (line 1) and input the sample data into the Sample Selection Embedding Module for sample differentiation and embedding (line 4). The resulting embeddings are then fed into the Differentiated FI Encoders to obtain different prediction outcomes (line 5). Next, the Dynamic Fusion Module integrates these outputs to generate the final prediction (line 6). At this point, a single prediction cycle is completed. Then, the Twin Focus Loss is computed to assist both modules and the encoder in adaptively differentiating the difficulty of the samples (line 7), while also calculating the CTR loss for the final prediction (line 8). The sum of these two losses constitutes the total loss (line 9), which is used to guide the model's optimization. Finally, we compute the average gradient for each mini-batch and update the model parameters based on the gradients (lines 11-12).

It is worth mentioning that Differentiated FI Encoders can be replaced by encoders from other models in line 5 of Algorithm \ref{training_process}. By utilizing the TF4CTR, Differentiated FI Encoders can be substituted with various other encoder architectures from different models, enabling flexible adaptation and optimization for different types of datasets and tasks. This modularity allows for fine-tuning of the model’s feature interaction capabilities, ensuring that the framework can be tailored to different recommendation or prediction tasks.

\subsection{Theoretical Analysis}
\label{Theoretical Analysis}
In most CTR models with parallel structures, a simple logical sum fusion method is widely favored \cite{deepfm,autoint,xdeepfm}, taking the form of $\hat{y} = \sigma(z_s+z_c)$. To keep the formula derivation simple and understandable, we will use this form of fusion as a basis for deriving the gradients obtained for different types of samples.
\subsubsection{Gradient of $\mathcal{L}_{ctr}$ for Encoders}
The gradient of $\mathcal{L}_{ctr}$ for logit values output by different encoders can be derived as:
\begin{equation}
\begin{aligned}
\|\nabla_{(z_s)} \mathcal{L}_{ctr}\| & = \frac{\partial \left(\log\sigma(z_s+z_c)\right)}{\partial z_s} = 1-\sigma\left(z_s+z_c\right), \\
\|\nabla_{(z_c)} \mathcal{L}_{ctr}\| & = \frac{\partial \left(\log\sigma(z_s+z_c)\right)}{\partial z_c} = 1-\sigma\left(z_s+z_c\right).
\end{aligned}
\label{ctr loss}
\end{equation}
This equation proves that $\mathcal{L}_{ctr}$ gives the same gradient for encoders with different task objectives. This often leads to untargeted learning of the encoders, thus weakening their performance and generalization ability.


\begin{figure}[t]
    \subfloat[$\mathcal{L}_{ctr} + \mathcal{L}_{TF}$]{
        \centering
        \includegraphics[width=0.48\linewidth]{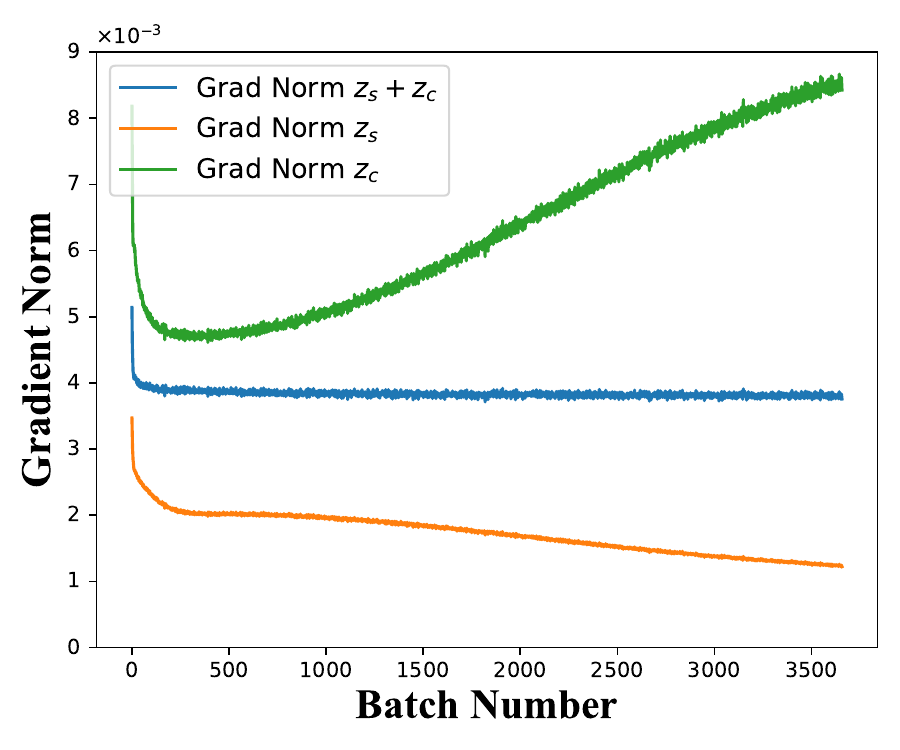}%
    }
    \subfloat[$\mathcal{L}_{ctr}$ only]{
        \centering
        \includegraphics[width=0.48\linewidth]{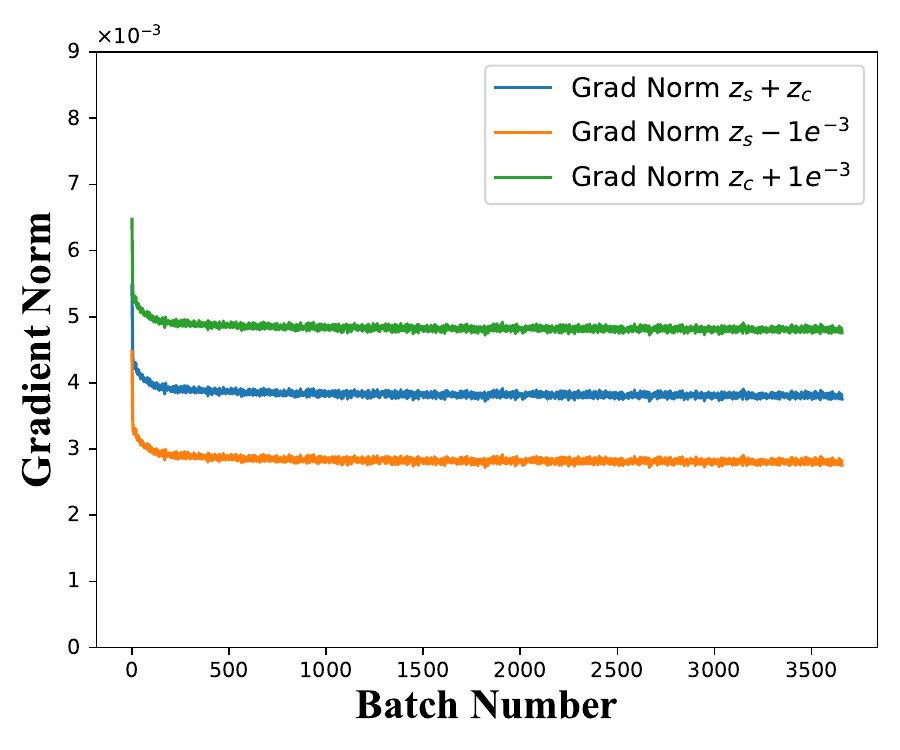}%
    }

    \captionsetup{justification=raggedright}
    \caption{Gradient norm dynamics of $\mathcal{L}_{ctr} + \mathcal{L}_{TF}$ and $\mathcal{L}_{ctr}$ for the first training epoch on the Criteo dataset.}
    \label{grad_norm}
\end{figure}

\subsubsection{Gradient of $\mathcal{L}_{TF}$ for Encoders}
\begin{equation}
\begin{aligned}
\|\nabla_{(z_s)} \mathcal{L}_{simple}\| & = \frac{\partial \left((c+\sigma(z_s))^\gamma\log\sigma(z_s)\right)}{\partial z_s} \\
&\propto (c+\sigma(z_s))^{\gamma-1}(1-\sigma(z_s)), \\
\|\nabla_{(z_c)} \mathcal{L}_{complex}\| & = \frac{\partial \left((m-\sigma(z_c))^\gamma\log\sigma(z_c)\right)}{\partial z_c} \\
&\propto (m-\sigma(z_c))^{\gamma-1}(1-\sigma(z_c)), \\
\end{aligned}
\end{equation}
From this equation, it can be seen that $\mathcal{L}_{TF}$ can be tuned to the gradient size accepted by different encoders using the hyperparameters $c$ and $\gamma$. For example, when the simple encoder encounters a hard sample (i.e., the gap between the predicted value and the true label is too large), the $\sigma(z_s)$ at this point tends to be a smaller value, which results in the value of $c+\sigma(z_s)$ being less than 1. The $\gamma$ amplifies this deflated scale, thus the loss results in the simple encoder accepts a lower gradient from difficult samples, and the opposite for complex encoders. On the other hand, as shown in Figure \ref{Unbalanced}, we weight the easy samples since they represent a larger proportion of the total samples, and the complete gradient of $\mathcal{L}_{TF}$ can be derived as:

{
\begin{equation}
\begin{aligned}
\|\nabla_{(z_s)} \mathcal{L}_{TF}\| & \propto \alpha (c+\sigma(z_s))^{\gamma-1}(1-\sigma(z_s)),\\
\|\nabla_{(z_c)} \mathcal{L}_{TF}\| & \propto (1-\alpha) (m-\sigma(z_c))^{\gamma-1}(1-\sigma(z_c)).
\end{aligned}
\end{equation}

Since $c\in(0,1)$, $m=2-c\in(1,2)$ and $\sigma(z_s),\sigma(z_c)\in(0,1)$, we have
$c+\sigma(z_s)\in(c,1+c)\subset(0,2)$ and $m-\sigma(z_c)\in(m-1,m)\subset(0,2)$.
Therefore the factors $(c+\sigma(z_s))^{\gamma-1}$ and $(m-\sigma(z_c))^{\gamma-1}$ are always bounded for $\gamma\in\{1,2,3\}$, and TF Loss only introduces a bounded, sample-adaptive rescaling of the standard cross-entropy gradients.
In particular, when a sample is hard (i.e., $\sigma(z)$ is small for a positive instance), $(c+\sigma(z_s))^{\gamma-1}<1$ and $(m-\sigma(z_c))^{\gamma-1}>1$ for $\gamma>1$, so the simple encoder receives a down-weighted gradient, while the complex encoder receives an amplified one; the opposite holds for easy samples.
This complementary behavior and the absence of gradient explosion/vanishing are consistent with the empirical gradient norm dynamics shown in Fig.~\ref{grad_norm}.
}

We aim to further demonstrate whether $\mathcal{L}_{TF}$ generates varying gradients for encoders, enabling them to precisely capture target-specific feature interaction information across different samples. We visualized the gradient norm obtained from logit values on the Criteo dataset, and the results are shown in Figure \ref{grad_norm}. In Figure \ref{grad_norm} (a), benefiting from the addition of $\mathcal{L}_{TF}$, the logit $z_c$ of the complex encoder output is gradually increasing, while the logit $z_s$ of the simple encoder is gradually decreasing. This implies that after the model has quickly fitted the easy samples, it gradually emphasizes capturing more effective information from the challenging and hard samples. However, as shown in Figure \ref{grad_norm} (b), $z_c$ and $z_s$ have no change in the gradient that they are subjected to as the number of batch increases. This implies that the model does not realize that it is only fitting the easy samples, which reduces the model performance.

\subsection{Complexity Analysis}
\subsubsection{Space Complexity.} In Section \ref{FI Encoder}, we introduce the Differentiated FI Encoder and validate our approach using simple MLPs of varying sizes. For ease of discussion, we denote $W_{\Psi}$ as the set of weight parameters in the corresponding MLP. Meanwhile, due to the introduction of the embedding layer, the corresponding space complexity is \textit{O}($dfs$). Since the parameters of the Sample Selection Embedding Module and Dynamic Fusion Module are typically much smaller than $W_{\Psi}$, we neglect them. Therefore, the space complexity of TF4CTR is approximately \textit{O}($dfs + 2|W_{\Psi}|$).

\subsubsection{Time Complexity.} Similar to the calculation of space complexity, the time complexity of $\mathcal{L}_{total}$ is \textit{O}(3$N$) due to the presence of three different loss functions. Since TF4CTR only uses simple MLPs as the FI Encoder, there is no explicit feature interaction computation overhead. Ultimately, we compute that the time complexity of TF4CTR during the training phase is approximately \textit{O}($dfs + 2|W_{\Psi}|$ + 3$N$), while the computational overhead associated with the loss functions is removed during the inference phase. More detailed results are shown in Table \ref{TimeComplexity}. To provide a clearer comparison of TF4CTR's time complexity, we also conducted an empirical comparison with several baseline models. The specific results can be found in Section \ref{Running Time Comparison}.

In addition, we present the training time complexities of DCNv2, xDeepFM, EDCN, AFN+, GDCN, and CL4CTR in Table \ref{TimeComplexity}. We let $L$, $U$, $M$, and $D^\prime$ represent the number of explicit interaction layers, logarithmic neurons of AFN+, feature maps of xDeepFM, and output dimension of the FI Encoder in CL4CTR, respectively. Overall, the number of parameters and computational cost of the model in TF4CTR are comparable to those of the baseline models. The computational cost of the loss functions is slightly higher but still much lower than that of CL4CTR. This ensures that the overall complexity of TF4CTR remains within a manageable range.

\begin{table}[t]
\renewcommand\arraystretch{1.2}
\centering
\caption{Comparison of Analytical Time Complexity \\
$s \gg  |W_{\Psi}| > D > f \approx d \approx M \approx U > L$}
\resizebox{1\linewidth}{!}{
\begin{tabular}{c|c|c|c|c}
\hline \textbf { Model } & \textbf { Embedding } & \textbf {Implicit interaction} & \textbf {Explicit interaction} & \textbf { Loss Function } \\
\hline \text { DCNv2 \cite{dcnv2}} & \textit{O}($dfs$) & \textit{O}($|W_{\Psi}|$) & \textit{O}($D^2L$) & \textit{O}($N$)\\
\hline \text { xDeepFM \cite{xdeepfm} } & \textit{O}($dfs$) & \textit{O}($|W_{\Psi}|$) & \textit{O}($dfM(1 + ML))$) & \textit{O}($N$)\\
\hline \text { EDCN \cite{EDCN}} & \textit{O}($dfs$) & \textit{O}($D^2L$) & \textit{O}($D^2L$) & \textit{O}($N$)\\
\hline \text { AFN+ \cite{AFN}} & \textit{O}(2$dfs$) & \textit{O}($|W_{\Psi}|$) & \textit{O}($2df(1 + U)$) & \textit{O}($N$)\\
\hline \text { GDCN \cite{GDCN}} & \textit{O}($dfs$) & \textit{O}($|W_{\Psi}|$) & \textit{O}($2D^2L$) & \textit{O}($N$)\\
\hline \text { CL4CTR \cite{CL4CTR}} & \textit{O}($dfs$) & \textit{O}(3$|W_{\Psi}|$) & - & \textit{O}($N + ND^\prime + \frac{dfN(1 + N)}{2} + \frac{dfN^2(f - 1)}{2}$)\\
\hline \text { TF4CTR } & \textit{O}($dfs$) & \textit{O}(2$|W_{\Psi}|$) & - & \textit{O}(3$N$)\\
\hline
\end{tabular}}
\label{TimeComplexity}
\end{table}

\section{Experiments}
In this section, we conduct extensive experiments on five real-world CTR prediction datasets to validate the effectiveness and compatibility of the proposed TF4CTR framework and address the following research questions (RQs): 
\begin{itemize} 
\item \textbf{RQ1} Can the proposed TF Loss be generalized across different models? Moreover, does it outperform similar loss functions? 
\item \textbf{RQ2} Do the proposed SSEM and DFM synergistically enhance model performance when combined? Which combination is the highest for performance improvement? 
\item \textbf{RQ3} Does our proposed model framework exhibit superiority over other models or frameworks?
\item \textbf{RQ4} Is the TF4CTR framework well compatible with other representative baseline models?
\item \textbf{RQ5} How should the hyperparameters of TF Loss be configured and balanced?
\item \textbf{RQ6} Is the TF4CTR framework lightweight and does it improve the generalization of the model?
\end{itemize}

\subsection{Experimental Settings}

\textbf{Datasets.} Table \ref{dataset}  provides detailed information about these datasets. Detailed descriptions of these datasets are in the references and links. We evaluate TF4CTR on five real-world datasets: 
\begin{itemize}
    \item Criteo\footnote{\url{https://www.kaggle.com/c/criteo-display-ad-challenge}} \cite{openbenchmark}: It is the well-known CTR benchmark dataset, which contains a 7-day stream of real data from Criteo, covering 39 anonymous feature fields.
    \item KKBox\footnote{\url{https://www.kkbox.com/intl}} \cite{Bars}: It aims to predict the chances of a user listening to a song repetitively after the first observable listening event within a time window was triggered. 
    \item ML-1M\footnote{\url{https://grouplens.org/datasets/movielens}} \cite{autoint}: It contains about 1 million user rating data for movies and is a standard dataset widely used in recommender system research.
    \item ML-tag\footnote{\url{https://github.com/reczoo/Datasets/tree/main/MovieLens}} \cite{AFN} \cite{Bars}: It consists of users' tagging records on movies. The datasets have been widely used in various research on recommender systems.
    \item Frappe\footnote{\url{http://baltrunas.info/research-menu/frappe}} \cite{AFN, frappe}: It contains app usage logs from users under different contexts (e.g., daytime, location). The target value indicates whether the user has used the app under the context.
\end{itemize}


\textbf{Data preprocessing and Feature Details.} Detailed statistics of the datasets, including the number of users, items, total features, and positive ratios, are summarized in Table \ref{dataset}. The specific feature lists and preprocessing methods applied are detailed in Table \ref{tab:feature_list}. For data preprocessing, we generally follow the protocols established by \cite{openbenchmark}\footnote{\url{https://github.com/reczoo/BARS/tree/main/datasets}}.

Regarding time-division boundaries, for datasets containing explicit timestamps (\textbf{ML-1M} and \textbf{KKBox}), we split the data chronologically. The \textbf{ML-1M} dataset covers interactions from May 2000 to February 2003, while \textbf{KKBox} spans from January 2004 to October 2017. For \textbf{Criteo}, \textbf{Frappe}, and \textbf{ML-tag}, we adopted a stratified random splitting strategy to maintain distribution consistency, as specific time boundaries are not applicable for these benchmark versions.
Notably, due to the Criteo dataset's large size and extremely high sparsity (over 99.99\%) \cite{autoint}, we implement a threshold of 10, replacing features occurring less than this threshold with a default "Out-of-Vocabulary" (OOV) token. For other datasets, a threshold of 2 is applied. Regarding the Criteo dataset, it is characterized by an extremely large feature space with high sparsity ($>99.99\%$). A higher threshold of 10 is widely adopted in the community to filter out long-tail noise (features appearing fewer than 10 times). This step is crucial for reducing the embedding table size to a manageable level without sacrificing significant predictive information. Specifically, the procedure involves a frequency-based filtration and re-indexing step. First, we compute the global occurrence count for every unique categorical feature value in the training set. Feature values with a frequency strictly lower than the threshold are identified as rare and are collectively mapped to a single, shared token, denoted as the OOV token. Conversely, feature values meeting or exceeding the threshold are retained.  Meanwhile, on the Criteo dataset, we discretize numerical features by rounding down each numeric value $x$ to $\lfloor \log^2(x) \rfloor$ for $x > 2$, and setting $x = 1$ otherwise \cite{openbenchmark}.


\begin{table*}[t]
{
\renewcommand\arraystretch{1.2}
\centering
\caption{Detailed Statistics of Datasets}
\label{dataset}
\resizebox{0.95\linewidth}{!}{
\begin{tabular}{c|c|c|c|c|c|c|c} 
\toprule 
\textbf{Dataset} & \textbf{\# Users} & \textbf{\# Items} & \textbf{\# Samples} & \textbf{\# Fields} & \textbf{\# Features} & \textbf{Pos. Ratio} & \textbf{Split} \\
\midrule 
\textbf{Criteo}  & N/A$^*$ & N/A$^*$ & 45,840,617 & 39 & 5,549,252 & 25.6\% & Random (8:1:1) \\
\textbf{KKBox}   & 30,755 & 359,966 & 7,377,418 & 19 & 227,853 & 50.4\% & Time-based (8:1:1) \\
\textbf{ML-1M}   & 6,040  & 3,706   & 1,000,209 & 5  & 9,776   & 57.5\% & Time-based (8:1:1) \\
\textbf{ML-tag}  & 17,045 & 23,743  & 2,006,859 & 3  & 88,596  & 33.3\% & Random (7:2:1) \\
\textbf{Frappe}  & 957    & 4,082   & 288,609   & 10 & 5,382   & 33.3\% & Random (7:2:1) \\
\bottomrule
\multicolumn{8}{l}{\footnotesize $^*$Criteo contains anonymous features; specific User/Item IDs are not explicitly defined. \# Features denotes the vocabulary size.} \\
\end{tabular}}
}
\end{table*}

\begin{table*}[t]
{
\renewcommand\arraystretch{1.2}
\centering
\caption{Feature Lists and Preprocessing Methods Used in Experiments}
\label{tab:feature_list}
\resizebox{0.95\linewidth}{!}{
\begin{tabular}{c|p{8cm}|p{6cm}}
\toprule
\textbf{Dataset} & \textbf{Feature List} & \textbf{Preprocessing / Encoding} \\
\midrule
\textbf{Criteo} & \textbf{Dense (13):} I1 -- I13; \textbf{Sparse (26):} C1 -- C26 & Log transformation \& Z-score (Dense); Label Encoding (Sparse). \\
\midrule
\textbf{KKBox} & msno, song\_id, source\_system\_tab, source\_screen\_name, source\_type, city, bd, gender, registered\_via, dates, song\_length, genre\_ids, artist\_name, composer, lyricist, language, name, isrc & Label Encoding (IDs/Text); Log transform (song\_length); Binning (bd); Multi-hot (genre\_ids). \\
\midrule
\textbf{ML-1M} & UserID, MovieID, Gender, Age, Occupation & Label Encoding + Embedding. \\
\midrule
\textbf{ML-tag} & user\_id, item\_id, tag\_id & Label Encoding + Embedding. \\
\midrule
\textbf{Frappe} & user, item, daytime, weekday, isweekend, homework, cost, weather, country, city & Label Encoding + Embedding. \\
\bottomrule
\end{tabular}}
}
\end{table*}

\begin{table*}[t]
\renewcommand\arraystretch{1.2}
\Huge
\centering
\caption{Compatibility study and performance comparison of TF Loss. $\triangle \text{AUC}$ denote the absolute performance improvement. AVG.AUC and AVG.RI denotes the average AUC performance and average absolute improvement on the four datasets, respectively. Typically, CTR researchers consider an improvement of \textbf{\textit{0.1\%}} in AUC to be statistically significant \cite{CL4CTR,openbenchmark}.} 
\label{loss_study}
\resizebox{\linewidth}{!}{
\begin{tabular}{c|c|c|cc|cc|cc|cc|c|c}
\hline \hline
\multirow{2}{*}{FI Encoder} & \multirow{2}{*}{Base Model} & \multirow{2}{*}{Loss function} & \multicolumn{2}{c|}{KKBox} & \multicolumn{2}{c|}{Criteo} & \multicolumn{2}{c|}{ML-tag} & \multicolumn{2}{c|}{Frappe} & \multirow{2}{*}{AVG.AUC} & \multirow{2}{*}{AVG.RI} \\ \cline{4-11}
                            &                             &                                & AUC(\%)$\uparrow$          & $\triangle \text{AUC}\uparrow$        & AUC(\%)$\uparrow$           & $\triangle \text{AUC}\uparrow$         & AUC(\%)$\uparrow$             & $\triangle \text{AUC}\uparrow$          & AUC(\%)$\uparrow$           & $\triangle \text{AUC}\uparrow$         &                          &                         \\ \hline
\multirow{4}{*}{MLP}        & \multirow{4}{*}{DualMLP}    & Log Loss (Base)                      & 84.08              & -           & 81.40              & -            & 96.96                & -             & \underline{98.37}              & -            & 90.20                         & -                        \\
                            &                             & Focal Loss \cite{focalloss}                    & 83.83              & -0.25           & \underline{81.42}              & +0.02            & 96.99                & +0.03             & 98.29              & -0.08            & 90.13                         & -0.07                        \\
                            &                             & R-CE Loss  \cite{rce}                     & \underline{84.20}              & +0.12           & 81.36              & -0.04            & \underline{97.07}                & +0.11             & 98.35              & -0.02            & \underline{90.24}                         & +0.04                        \\
                            &                             & \textbf{TF Loss (Ours)}                 & \textbf{84.29}              & +0.21           & \textbf{81.48}              & +0.08            & \textbf{97.12}                & +0.16             & \textbf{98.55}              & +0.18            & \textbf{90.36}                         & \textbf{+0.16}                        \\ \hline
\multirow{4}{*}{Attention}  & \multirow{4}{*}{AutoInt+}   & Log Loss (Base)                      & \underline{84.08}         & -           & 81.39         &  -           & 96.97           &  -            & 98.41         & -            & 90.21                        & -                        \\
                            &                             & Focal Loss \cite{focalloss}                    & 83.91              & -0.17           & 81.35              & -0.04            & 96.76                & -0.21             & \underline{98.53}              & +0.12            & 90.13                         & -0.08                        \\
                            &                             & R-CE Loss  \cite{rce}                     & 84.01              & -0.07           & \underline{81.40}               & +0.01            & \underline{97.08}                & +0.11             & 98.46              & +0.05            & \underline{90.23}                         & +0.02                        \\
                            &                             & \textbf{TF Loss (Ours)}                 & \textbf{84.25}              & +0.17            & \textbf{81.46}              & +0.07            & \textbf{97.11}                & +0.14             & \textbf{98.56}              & +0.15            & \textbf{90.34}                         & \textbf{+0.13}                        \\ \hline
\multirow{16}{*}{Product}   & \multirow{4}{*}{DeepFM}     & Log Loss (Base)                       & \underline{84.11}         & -           & 81.38         & -            & 96.92           & -             & 98.37         & -            & 90.19                         & -                        \\
                            &                             & Focal Loss \cite{focalloss}                    & 83.81             & -0.30           & 81.33              & -0.05            & \underline{96.95}                & +0.03             & \underline{98.43}              & +0.06            & 90.13                         & -0.06                        \\
                            &                             & R-CE Loss  \cite{rce}                     & \underline{84.11}              & +0.00           & \underline{81.41}              & +0.03            & 96.92                & +0.00             & 98.42              & +0.05            & \underline{90.21}                         & +0.02                        \\
                            &                             & \textbf{TF Loss (Ours)}                 & \textbf{84.31}              & +0.20           & \textbf{81.43}              & +0.05            & \textbf{96.99}                & +0.07             & \textbf{98.48}              & +0.11            & \textbf{90.30}                         & \textbf{+0.11}                        \\ \cline{2-13} 
                            & \multirow{4}{*}{DCN}        & Log Loss (Base)                      & \underline{84.22}         & -           & \underline{81.39}         & -            & 96.91           & -             & 98.38         & -            & \underline{90.22}                         & -                        \\
                            &                             & Focal Loss \cite{focalloss}                    & 84.07              & -0.15           & 81.38              & -0.01            & 96.85                & -0.06             & \underline{98.46}              & +0.08            & 90.19                         & -0.03                        \\
                            &                             & R-CE Loss  \cite{rce}                     & 84.00              & -0.22           & 81.32              & -0.07            & \underline{97.10}                & +0.19             & 98.39              & +0.01            & 90.20                         & +0.02                        \\
                            &                             & \textbf{TF Loss (Ours)}                 & \textbf{84.35}              & +0.13           & \textbf{81.44}              & +0.05            & \textbf{97.12}                & +0.21             & \textbf{98.53}              & +0.15            & \textbf{90.36}                         & \textbf{+0.14}                        \\ \cline{2-13} 
                            & \multirow{4}{*}{DCNv2}      & Log Loss (Base)                      & \underline{84.24}         & -           & 81.40         & -            & 96.92           & -             & 98.45         & -            & \underline{90.25}                         & -                        \\
                            &                             & Focal Loss \cite{focalloss}                    & 83.86              & -0.38           & \underline{81.42}              & +0.02            & \underline{96.99}                & +0.07             & \underline{98.50}              & +0.05            & 90.19                         & -0.06                        \\
                            &                             & R-CE Loss  \cite{rce}                     & 84.18              & -0.06           & 81.40              & +0.00            & 96.90                & -0.02             & 98.46              & +0.01            & 90.23                         & -0.02                        \\
                            &                             & \textbf{TF Loss (Ours)}                 & \textbf{84.32}              & +0.08           & \textbf{81.45}              & +0.05            & \textbf{97.03}                & +0.11             & \textbf{98.52}              & +0.07            & \textbf{90.33}                         & \textbf{+0.08}                        \\ \cline{2-13} 
                            & \multirow{4}{*}{AFN+}       & Log Loss (Base)                      & 83.57         & -           & \underline{81.31}         & -            & 96.84           & -             & 98.19         & -            & 89.97                         & -                        \\
                            &                             & Focal Loss \cite{focalloss}                    & 82.92              & -0.65           & 81.25              & -0.06            & 96.43                & -0.41             & 97.79              & -0.40            & 89.59                         & -0.38                        \\
                            &                             & R-CE Loss  \cite{rce}                     & \underline{83.74}              & +0.17           & \underline{81.31}              & +0.00            & \underline{96.88}                & +0.04             & \underline{98.20}              & +0.01            & \underline{90.03}                         & +0.06                        \\
                            &                             & \textbf{TF Loss (Ours)}                 & \textbf{84.37}              & +0.80           & \textbf{81.42}              & +0.11            & \textbf{97.02}                & +0.18             & \textbf{98.45}              & +0.26            & \textbf{90.31}                         & \textbf{+0.34}                        \\ \hline \hline
\end{tabular}}
\end{table*}

\noindent\textbf{Evaluation metrics.} To compare the performance, we utilize two commonly used metrics in CTR models: \textbf{AUC}, \textbf{gAUC} \cite{autoint, GDCN, ComboFashion}. AUC stands for Area Under the ROC Curve, which measures the probability that a positive instance will be ranked higher than a randomly chosen negative one. gAUC implements user-level AUC computation.

\noindent\textbf{Baselines.} We compared TF4CTR with some state-of-the-art (SOTA) models (\textbf{+} denotes Integrating the original model with DNN networks): 
\begin{itemize}
    \item Focal Loss \cite{focalloss} (2017): it addresses class imbalance by down-weighting easy examples and focusing on hard ones, improving performance in imbalanced classification tasks.
    \item R-CE Loss \cite{rce} (2021): it enhances cross-entropy loss by applying a dynamic re-weighting mechanism, focusing on difficult samples to reduce the impact of easy ones in class-imbalanced scenarios.
    \item Wide\ \&\ Deep \cite{widedeep} (2016): a hybrid model combining a wide linear model and a deep neural network, aimed at leveraging both memorization and generalization for CTR prediction.
    \item DeepFM \cite{deepfm} (2017): it integrates the factorization machine with a deep neural network, efficiently modeling both feature interactions and non-linear patterns for CTR prediction tasks.
    \item DCN \cite{dcn} (2017): it introduces cross-network layers to explicitly model feature interactions, improving CTR prediction tasks.
    \item xDeepFM \cite{xdeepfm} (2018): it combines the strengths of DeepFM and a novel compressed interaction network to improve both feature interaction modeling and deep learning performance in CTR prediction tasks.
    \item AutoInt+ \cite{autoint} (2019): it extends the AutoInt model by enhancing self-attention mechanisms for automatic feature interaction learning, boosting performance in CTR prediction tasks.
    \item AFN+ \cite{AFN} (2020): it utilizes a logarithmic transformation layer to learn adaptive-order feature interactions, boosting performance in CTR prediction tasks.
    \item DCNv2 \cite{dcnv2} (2021): it enhances the original DCN model with a more efficient cross-network structure, boosting interaction learning and performance in large-scale recommender systems.
    \item EDCN \cite{EDCN} (2021): it improves DCN by introducing a more flexible bridge structure and applying it to CTR prediction tasks.
    \item CL4CTR \cite{CL4CTR} (2023): it focuses on contrastive learning for click-through rate prediction, leveraging both user-item interactions and implicit feedback to enhance model robustness.
    \item EulerNet \cite{EulerNet} (2023): it uses Euler's theorem-based feature interaction modeling, improving the learning of complex interactions in recommendation systems while maintaining computational efficiency.
    \item DLF \cite{DLF} (2025): it introduces a dynamic low-order-aware fusion framework that effectively integrates explicit and implicit feature interactions to enhance CTR prediction.
    
\end{itemize}

\noindent\textbf{Implementation Details.} We implement all models using PyTorch \cite{PYTORCH}, with all parameters initialized using the Xavier method. Regarding the distribution of model parameters and computations, we follow the organization of MMoE \cite{mmoe}, where both expert and gating networks are implemented via MLP. Specifically, the expert networks are configured with default hidden units of [400, 200], while the gating networks utilize [400, 100] units to control the contribution of each expert. We employ the Adam optimizer \cite{adam} to optimize all models, with a default learning rate set to 0.001. For the sake of fair comparison, we set the embedding dimension to 16 \cite{openbenchmark, Bars}, and the numbers of MLP hidden units for the main deep layers are [400, 400, 400] \cite{GDCN, EulerNet}. The batch size is set to 4,096 on the Criteo dataset and 10,000 on the other datasets. To prevent overfitting, we employ early stopping with a patience value of 2. The hyperparameters of the baseline model are configured and finetuned based on the \textit{optimal values} provided in \cite{FuxiCTR,openbenchmark} and their original paper. Further details on model hyperparameters and dataset configurations are available in our straightforward and accessible running logs\footnote{\url{https://github.com/salmon1802/TF4CTR/tree/main/TF4CTR/TF4CTR_torch/checkpoints}}.


\subsection{Effectiveness Analysis}
\subsubsection{Comparison and Compatibility Study of Loss (RQ1)}
To verify whether our proposed $\mathcal{L}_{TF}$ can be generalized to other SOTA models and outperform other loss functions, we fix the hyperparameters of the base model and conduct compatibility and comparison experiments. The results are shown in Table \ref{loss_study}. We bold the best performance, while underlined scores are the second best.

We can find that TF Loss gains performance on all four datasets relative to the base Logloss. This proves that the model optimized by TF Loss has better generalization ability and performance. It is worth mentioning that the AFN+ model does not perform well with Logloss optimization, but after optimization with TF Loss, the model achieves the same benchmarks as the other SOTA models, and even achieves the best performance on the KKBox dataset. Additionally, the design idea of TF Loss is to help encoders target-specific handling samples of varying difficulty, thereby enabling more precise adjustment of model parameters. 

Meanwhile, we further compared the performance differences when optimizing models with TF Loss and other loss functions. R-CE Loss often achieves the sub-optimal performance, while Focal Loss is not always effective, and often even degrades the performance of the model. We empirically think that this is due to the loss weights of Focal Loss always being less than 1, leading to a less effective supervision signal received by the encoders. Regarding R-CE Loss, it abandons the optimization of hard samples, thereby allowing the model to focus more on the information within challenging and easy samples. However, this approach is not always effective. For instance, in DCNv2, R-CE Loss led to negative optimization and recent work \cite{rankandlog} suggests that encouraging the model to learn simpler samples might lead to gradient vanishing issues. For TF Loss, its effectiveness in enhancing model performance across various datasets and models further confirms the validity of our approach.

\begin{table*}[t]
\centering
\caption{Performance comparison of different SSEM and DFM methods. Bold indicates the best-combined performance. The "Share" row can be viewed as an ablation study removing the SSEM, while the "Sum" column represents an ablation study removing the DFM.} 
\renewcommand{\arraystretch}{0.9}
\resizebox{\linewidth}{!}{
\begin{tabular}{c|c|cc|cc|cc|cc|cc}
\hline\hline
\multirow{6}{*}{ML-tag} &
  \multirow{2}{*}{Method} &
  \multicolumn{2}{c|}{WSF} &
  \multicolumn{2}{c|}{VF} &
  \multicolumn{2}{c|}{CF} &
  \multicolumn{2}{c|}{MoEF} &
  \multicolumn{2}{c}{Sum} \\ \cline{3-12} 
 &
   &
  AUC &
  gAUC &
  AUC &
  gAUC &
  AUC &
  gAUC &
  AUC &
  gAUC &
  AUC &
  gAUC \\ \cline{2-12} 
 &
  SER &
  \textit{\textbf{97.46}} &
  \textit{\textbf{96.50}} &
  96.96 &
  96.02 &
  97.25 &
  96.33 &
  97.23 &
  96.26 &
  97.44 &
  \textit{\textbf{96.50}} \\
 &
  GM &
  97.02 &
  96.07 &
  97.02 &
  96.09 &
  97.02 &
  96.07 &
  96.99 &
  95.99 &
  97.09 &
  96.19 \\
 &
  MoE &
  96.79 &
  95.72 &
  96.95 &
  95.94 &
  96.82 &
  95.79 &
  96.77 &
  95.72 &
  96.79 &
  95.77 \\
 &
  Share &
  97.06 &
  96.18 &
  97.04 &
  96.12 &
  97.03 &
  96.05 &
  97.00 &
  95.98 &
  97.11 &
  96.19 \\ \hline
\multirow{6}{*}{KKBox} &
  \multirow{2}{*}{Method} &
  \multicolumn{2}{c|}{WSF} &
  \multicolumn{2}{c|}{VF} &
  \multicolumn{2}{c|}{CF} &
  \multicolumn{2}{c|}{MoEF} &
  \multicolumn{2}{c}{Sum} \\ \cline{3-12} 
 &
   &
  AUC &
  gAUC &
  AUC &
  gAUC &
  AUC &
  gAUC &
  AUC &
  gAUC &
  AUC &
  gAUC \\ \cline{2-12} 
 &
  SER &
  84.47 &
  77.60 &
  84.17 &
  77.11 &
  84.36 &
  77.48 &
  \textit{\textbf{84.50}} &
  \textit{\textbf{77.62}} &
  84.41 &
  77.54 \\
 &
  GM &
  83.92 &
  76.82 &
  83.78 &
  76.60 &
  83.98 &
  76.95 &
  83.97 &
  76.88 &
  83.97 &
  77.00 \\
 &
  MoE &
  83.71 &
  76.53 &
  83.59 &
  76.27 &
  83.73 &
  76.51 &
  83.75 &
  76.63 &
  83.76 &
  76.52 \\
 &
  Share &
  84.05 &
  77.03 &
  83.83 &
  76.66 &
  84.11 &
  77.07 &
  84.08 &
  77.07 &
  84.09 &
  77.17 \\ \hline
\multirow{6}{*}{Frappe} &
  \multirow{2}{*}{Method} &
  \multicolumn{2}{c|}{WSF} &
  \multicolumn{2}{c|}{VF} &
  \multicolumn{2}{c|}{CF} &
  \multicolumn{2}{c|}{MoEF} &
  \multicolumn{2}{c}{Sum} \\ \cline{3-12} 
 &
   &
  AUC &
  gAUC &
  AUC &
  gAUC &
  AUC &
  gAUC &
  AUC &
  gAUC &
  AUC &
  gAUC \\ \cline{2-12} 
 &
  SER &
  \textit{\textbf{98.72}} &
  98.17 &
  98.52 &
  98.02 &
  98.70 &
  \textit{\textbf{98.21}} &
  98.64 &
  98.20 &
  \textit{\textbf{98.72}} &
  \textit{\textbf{98.21}} \\
 &
  GM &
  98.57 &
  98.06 &
  98.40 &
  97.88 &
  98.41 &
  97.84 &
  98.42 &
  97.92 &
  98.57 &
  97.97 \\
 &
  MoE &
  98.54 &
  98.00 &
  98.53 &
  98.00 &
  98.56 &
  98.01 &
  98.44 &
  97.91 &
  98.55 &
  97.97 \\
 &
  Share &
  98.52 &
  98.02 &
  98.51 &
  97.98 &
  98.54 &
  98.04 &
  98.41 &
  97.85 &
  98.51 &
  97.98 \\ \hline
\multirow{6}{*}{ML-1M} &
  \multirow{2}{*}{Method} &
  \multicolumn{2}{c|}{WSF} &
  \multicolumn{2}{c|}{VF} &
  \multicolumn{2}{c|}{CF} &
  \multicolumn{2}{c|}{MoEF} &
  \multicolumn{2}{c}{Sum} \\ \cline{3-12} 
 &
   &
  AUC &
  gAUC &
  AUC &
  gAUC &
  AUC &
  gAUC &
  AUC &
  gAUC &
  AUC &
  gAUC \\ \cline{2-12} 
 &
  SER &
  \textit{\textbf{81.94}} &
  \textit{\textbf{77.50}} &
  81.57 &
  77.01 &
  81.32 &
  76.79 &
  81.46 &
  76.73 &
  81.83 &
  77.39 \\
 &
  GM &
  81.61 &
  76.88 &
  81.27 &
  76.45 &
  81.36 &
  77.01 &
  81.24 &
  76.58 &
  81.41 &
  76.76 \\
 &
  MoE &
  81.27 &
  76.56 &
  81.26 &
  76.42 &
  81.29 &
  76.55 &
  81.31 &
  76.74 &
  81.19 &
  76.39 \\
 &
  Share &
  81.58 &
  77.05 &
  81.33 &
  76.52 &
  81.41 &
  77.00 &
  81.35 &
  76.95 &
  81.26 &
  76.69 \\ \hline\hline
\end{tabular}}
\label{SSEMDFM}
\end{table*}

\begin{table*}[t]

\renewcommand\arraystretch{1.2}
\Huge
\centering
\caption{Performance comparison of different models. We conducted a two-tailed T-test ($p$-values) to assess the statistical significance between the TF4CTR and the best baseline model. It is worth noting that even a slight improvement (e.g., \textbf{0.1\%}) in AUC is meaningful in the context of CTR prediction tasks \cite{EDCN,CL4CTR,openbenchmark}.} 
\resizebox{\linewidth}{!}{
\begin{tabular}{c|c|ccccccccccc|cc}
\hline\hline \multirow{1}{*}{ Dataset } & Metric & WideDeep & DeepFM & DCN & xDeepFM & AutoInt+ & AFN+ & DCNv2 & EDCN & CL4CTR & EulerNet & DLF & TF4CTR & $p$-values\\
\hline \multirow{1}{*}{ KKBox } & $\text{AUC(\%)}\uparrow$ & 84.12 & 84.11 & 84.22 & 84.02 & 84.08 & 83.57 & \underline{84.24} & 83.96 & 84.05 & 83.68 & 84.07 & $\mathbf{84.50}$ & 2.44e-4\\
\hline \multirow{1}{*}{ ML-1M } & $\text{AUC(\%)}\uparrow$ & 81.22 & 81.37 & 81.38 & 81.37 & 81.30 & 81.37 & 81.44 & 81.22 & 81.13 & \underline{81.45}  & 81.44 & $\mathbf{81.94}$ & 4.79e-4\\
\hline \multirow{1}{*}{ ML-tag } & $\text{AUC(\%)}\uparrow$ & 96.92 & 96.92 & 96.91 & 96.92 & \underline{96.97} & 96.84 & 96.92 & 96.03 & 96.83 & 96.79 & 96.94 & $\mathbf{97.46}$ & 1.49e-4\\
\hline \multirow{1}{*}{ Frappe } & $\text{AUC(\%)}\uparrow$ & 98.32 & 98.37 & 98.38 & \underline{98.45} & 98.41 & 98.19 & \underline{98.45} & 98.41 & 98.27 & 98.16  & 98.40 & $\mathbf{98.72}$ & 7.26e-5 \\
\hline \multirow{1}{*}{ Criteo } & $\text{AUC(\%)}\uparrow$ & 81.35 & 81.38 & 81.39 & \underline{81.40} & 81.39 & 81.31 & \underline{81.40} & 81.39 & 81.36 & 81.37  & \underline{81.40} & $\mathbf{81.50}$ & 4.29e-5 \\
\hline\hline
\end{tabular}} \newline
\label{baseline}
\end{table*}

\subsubsection{Combination and Ablation Study of SSEM and DFM Method (RQ2)}
To validate which combination of SSEM and DFM is more effective, we conduct experiments with a fixed $\alpha=0.25$ and a 3-layer $\texttt{MLP}_c$, as shown in Table \ref{SSEMDFM}. Meanwhile, in order to provide a clearer comparison of the performance gains brought by SSEM and DFM, we have added the "Share" row and "Sum" column.
\begin{itemize}
    \item \textbf{Share} means that instead of using SSEM, different encoders are allowed to share the same feature embedding.
    \item \textbf{Sum} means that the logit values output by the encoders are subjected to a simple summing operation to obtain the final model prediction.
\end{itemize}
We can draw several findings from the results:
\begin{itemize} 
\item The SER method demonstrated superior performance across all four datasets. This proves the validity of the SSEM model. Empirically, we attribute this to its ability to further decouple the embedding representation learning process of samples, allowing for gradients to propagate back without mutual interference.

\item Sophisticated fusion methodologies do not guarantee superior effectiveness; conversely, the simpler WSF and Sum methods can also yield enhanced performance. For instance, the WSF approach demonstrated exceptional results on the ML-tag, Frappe, and ML-1M datasets, substantiating the effectiveness of the DFM. Notably, WSF introduces a minimal computational overhead with only two additional parameters compared to the Sum method, and yet, it achieves a performance increment of \textbf{0.1\%} on the ML-1M dataset.

\item Certain integrations of SSEM and DFM yielded inferior results compared to the Share+Sum approach, exemplified by the MoE+VF combination on the ML-tag dataset and the GM+CF pairing on the KKBox dataset. However, on the Frappe and ML-1M datasets, these same combinations exhibited improvements over Share+Sum. We empirically consider that differences in the data distributions are the primary cause of this discrepancy. ML-tag and KKBox datasets are larger with more severe data distribution bias \cite{bias}, whereas Frappe and ML-1M have smaller data volumes and simpler data distributions.

\item The majority of SSEM+DFM pairings surpass the performance of Share+Sum. The optimal SSEM and DFM implementations outperform Share+Sum, achieving gains of 0.35\% in AUC and 0.31\% in gAUC on ML-tag, 0.21\% in AUC and 0.19\% in gAUC on Frappe, 0.41\% in AUC and 0.45\% in gAUC on KKBox, and 0.68\% in AUC and 0.81\% in gAUC on ML-1M. These results prove the effectiveness of our proposed SSEM and DFM.
\end{itemize}

\subsubsection{Performance Comparison with Other Models (RQ3)}
\label{othermodel}
To further validate the outstanding performance of our proposed TF4CTR, we compare it against ten representative baseline models. Bold and underlined numbers respectively indicate the best and second-best results. Additionally, we conduct five runs of TF4CTR and the second-best baseline, and calculate the corresponding $p$-values. Notably, since the goal of this paper is to propose a general CTR framework, we implement the FI Encoder using only a simple MLP. It is foreseeable that the performance of TF4CTR could be further improved by using more advanced encoders. Based on the results as shown in Table \ref{baseline}, we can make the following observations:

\begin{itemize} 
\item In smaller datasets such as ML-1M and Frappe, the performance gap among the ten selected representative baseline models is considerable, reaching up to 0.3\%. However, in the larger Criteo dataset, this gap narrows. We empirically believe that the larger scale of data typically aids models in more consistently learning the underlying data distribution, which reduces the variance in performance between models. With large datasets, even simple models are capable of learning complex feature representations due to the extensive data available. In contrast, within smaller datasets, the complexity of the models and their regularization strategies may significantly influence performance. Therefore, model selection and tuning strategies should be appropriately adapted to the scale of the dataset to achieve optimal performance.

\item Upon analyzing Tables \ref{loss_study} and \ref{baseline}, it is evident that TF4CTR consistently exhibits significant performance improvements over DualMLP. TF4CTR achieves an average AUC of 90.54\% across the KKBOX, Criteo, ML-tag, and Frappe datasets. Compared to DualMLP optimized with Log Loss, TF4CTR realizes an absolute enhancement of 0.34\%, affirming the overall effectiveness of our proposed framework. Furthermore, even against DualMLP optimized with TF Loss, our model still achieves an additional absolute gain of 0.18\%. This further validates the effectiveness of our proposed SSEM and DFM, which help the model to better adapt to the auxiliary supervision signals provided by TF Loss, thereby enhancing the model's performance and generalization capabilities.

\item TF4CTR consistently achieves superior performance across five datasets. Compared to the second-best model on each dataset, it secures performance gains of 0.26\% on KKBox, 0.49\% on ML-1M, 0.49\% on ML-tag, 0.27\% on Frappe, and 0.1\% on Criteo, all exceeding the significant improvement benchmark of \textbf{0.1\%}. Furthermore, two-tailed T-tests comparing TF4CTR with the second-best models yield $p$-values below 0.01 \cite{GDCN}, indicating that the performance enhancements offered by our model are statistically significant.
\end{itemize}

\subsubsection{Compatibility Study of TF4CTR (RQ4)}
The TF4CTR framework proposed serves as a model-agnostic framework designed to enhance the performance of CTR models. To verify the compatibility of TF4CTR, it has been implemented into three benchmark baseline models, with experimental results detailed in Table \ref{Compatibility_TF4CTR}. The framework consistently enhances performance across multiple datasets. Notably, the WideDeep model achieves a performance increase of 0.41\% on the ML-1M dataset, elevating its rank from the second to last to the top-2 position in Table \ref{baseline}. In addition, both DeepFM and xDeepFM models exhibited performance improvements, with most $\triangle \text{AUC}$ values ex\textbf{}ceeding 0.1\%. These results substantiate the compatibility and effectiveness of the TF4CTR framework, underscoring its capacity to boost model performance across a spectrum of data environments consistently.

\begin{table}[t]
\renewcommand\arraystretch{1.0}
\centering
\caption{Compatibility study of TF4CTR.} 
\resizebox{1\linewidth}{!}{
\begin{tabular}{c|cc|cc|cc}
\hline\hline
\multirow{2}{*}{Model} & \multicolumn{2}{c|}{ML-tag} & \multicolumn{2}{c|}{Frappe} & \multicolumn{2}{c}{ML-1M} \\ \cline{2-7} 
                       & $\text{AUC(\%)}\uparrow$       & $\triangle \text{AUC}\uparrow$         & $\text{AUC(\%)}\uparrow$       & $\triangle \text{AUC}\uparrow$         & $\text{AUC(\%)}\uparrow$      & $\triangle \text{AUC}\uparrow$        \\ \hline
DeepFM                 & 96.92         & -           & 98.37         & -           & 81.37        & -          \\
$\text{DeepFM}_{\text{TF4CTR}}$     & \textbf{97.03}         & +0.11\%      & \textbf{98.54}         & +0.17\%      & \textbf{81.56}        & +0.19\%     \\ \hline
WideDeep               & 96.92         & -           & 98.32         & -           & 81.22        & -          \\
$\text{WideDeep}_{\text{TF4CTR}}$   & \textbf{97.04}         & +0.12\%      & \textbf{98.49}         & +0.17\%      & \textbf{81.63}        & +0.41\%     \\ \hline
xDeepFM                & 96.92         & -           & 98.45         & -           & 81.37        & -          \\
$\text{xDeepFM}_{\text{TF4CTR}}$    & \textbf{97.03}         & +0.11\%      & \textbf{98.56}         & +0.11\%      & \textbf{81.65}        & +0.28\%     \\ \hline \hline
\end{tabular}}
\label{Compatibility_TF4CTR}
\end{table}

\subsection{Hyper-parameter Analysis (RQ5)}
In this section, we further investigate the effects that the three important parameters of TF Loss bring to the model. Here we fix the experimental setup of Section \ref{othermodel} and change only the parameter to be studied.

\subsubsection{Impact of $\alpha$}
\begin{figure}[t]
    \subfloat{
        \centering
        \includegraphics[width=0.32\linewidth]{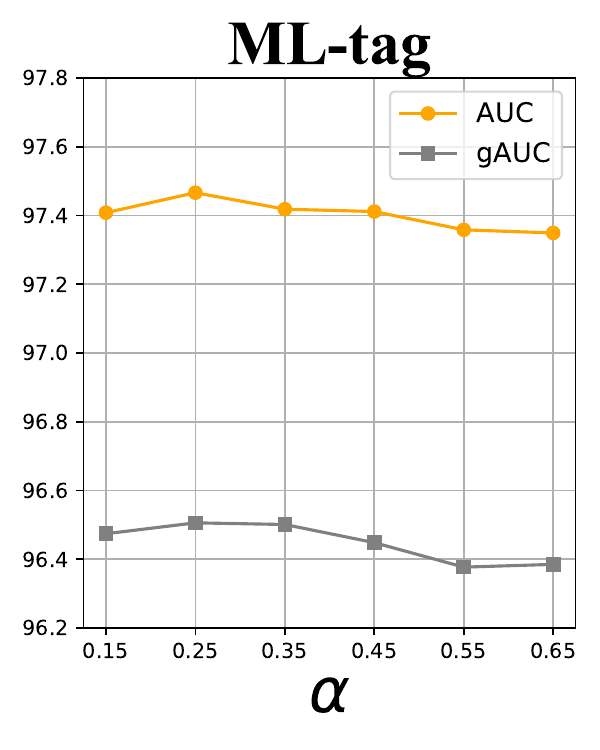}%
    }
    \subfloat{
        \centering
        \includegraphics[width=0.32\linewidth]{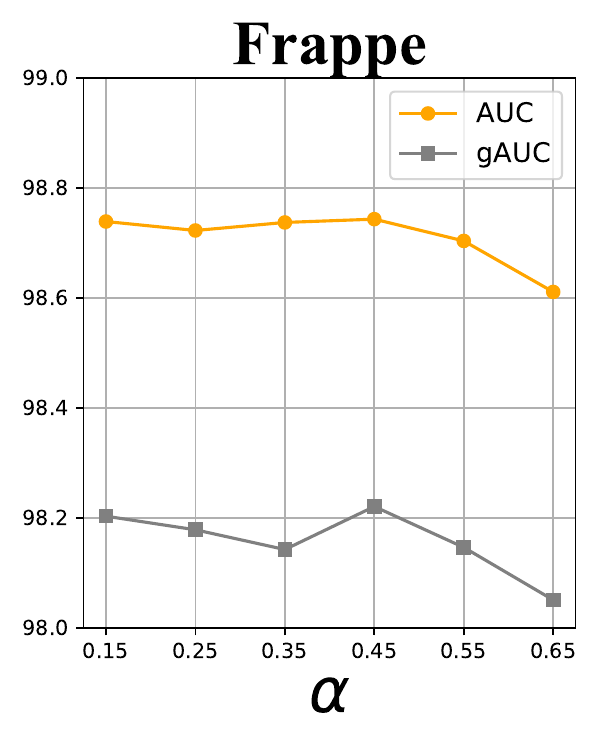}
    }
    \subfloat{
        \centering
        \includegraphics[width=0.32\linewidth]{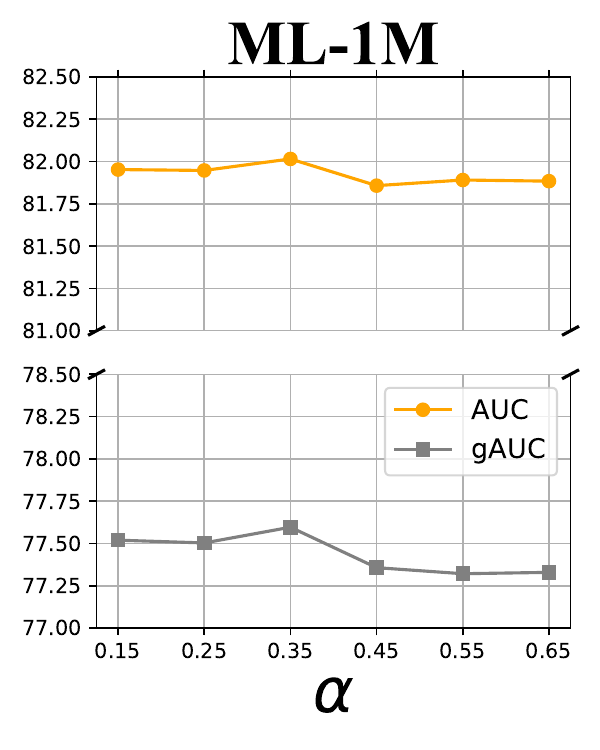}
    }
    \captionsetup{justification=raggedright}
    \caption{Influence of magnitude $\alpha$ of $\mathcal{L}_{TF}$.}
    \label{alpha}
\end{figure}

We fix the other hyperparameters and adjust $\alpha$ in the range [0.15, 0.65] in steps of 0.1, and the experimental results are shown in Figure \ref{alpha}. It can be observed that the model performance has a tendency to decrease as $\alpha$ increases and it reaches its optimal value at $\alpha=0.25$ on ML-tag, $\alpha=0.45$ on Frappe, and $\alpha=0.35$ on ML-1M. This further supports the idea that easy samples have a heavier portion in the dataset. We can make $\mathcal{L}_{TF}$ more adaptable to the sample distribution in the dataset by fine-grained adjustment of $\alpha$.

\subsubsection{Impact of $c$}
\begin{figure}[t]
    \subfloat{
        \centering
        \includegraphics[width=0.32\linewidth]{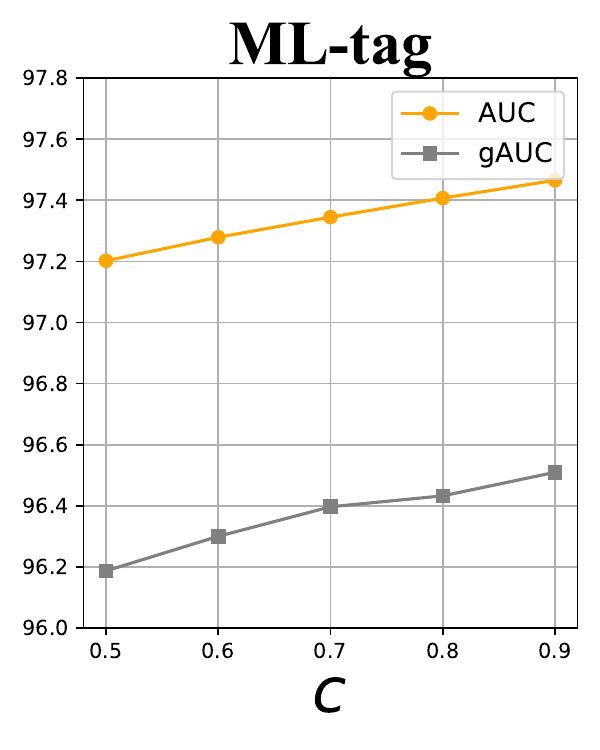}
    }
    \subfloat{
        \centering
        \includegraphics[width=0.32\linewidth]{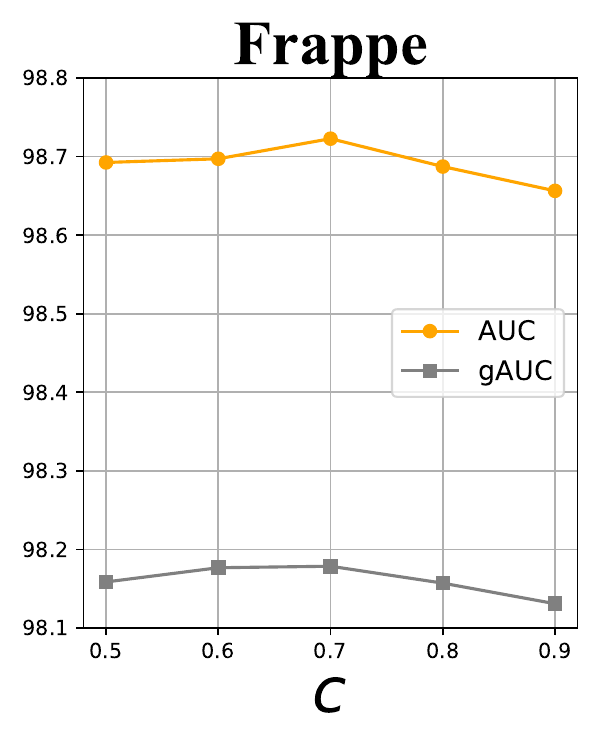}
    }
    \subfloat{
        \centering
        \includegraphics[width=0.32\linewidth]{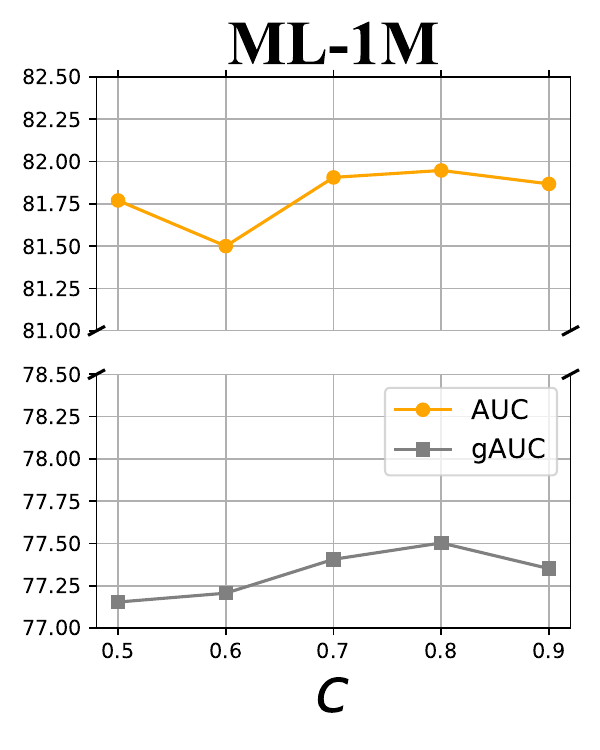}
    }
    \captionsetup{justification=raggedright}
    \caption{Influence of magnitude $c$ of $\mathcal{L}_{TF}$.}
    \label{impact_of_c}
\end{figure}

We introduce $c$ for $\mathcal{L}_{TF}$ aiming to use it as a threshold to classify \#\textit{Misclassified}, \#\textit{Poorly Classified}, and \#\textit{Well-classified}, thus further increasing or reducing the corresponding sample loss. The experimental results are shown in Figure \ref{impact_of_c}, where the model performance on ML-tag increases with $c$, which indicates a narrower distribution of difficult samples on ML-tag. On Frappe and ML-1M, the peak is reached when $c$ is 0.7 and 0.8, and then starts to decrease. This suggests that the distribution of difficult samples is not the same in different datasets, and we can further adjust $c$ to maximize the gain of $\mathcal{L}_{TF}$ for performance.

\subsubsection{Impact of $\gamma$}
\begin{figure}[t]
    \subfloat{
        \centering
        \includegraphics[width=0.32\linewidth]{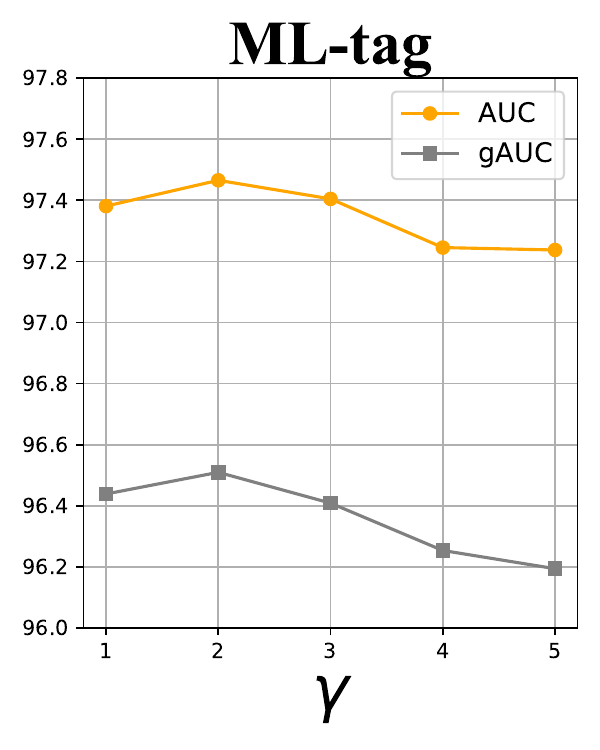}
    }
    \subfloat{
        \centering
        \includegraphics[width=0.32\linewidth]{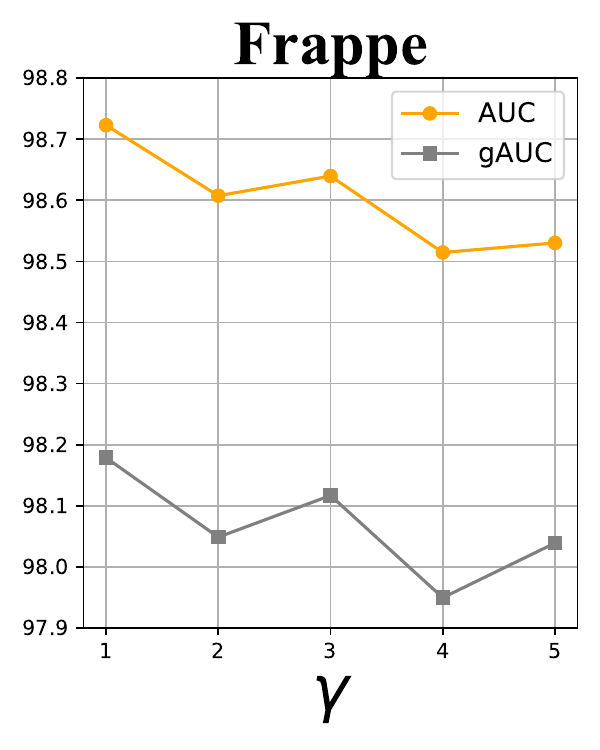}
    }
    \subfloat{
        \centering
        \includegraphics[width=0.32\linewidth]{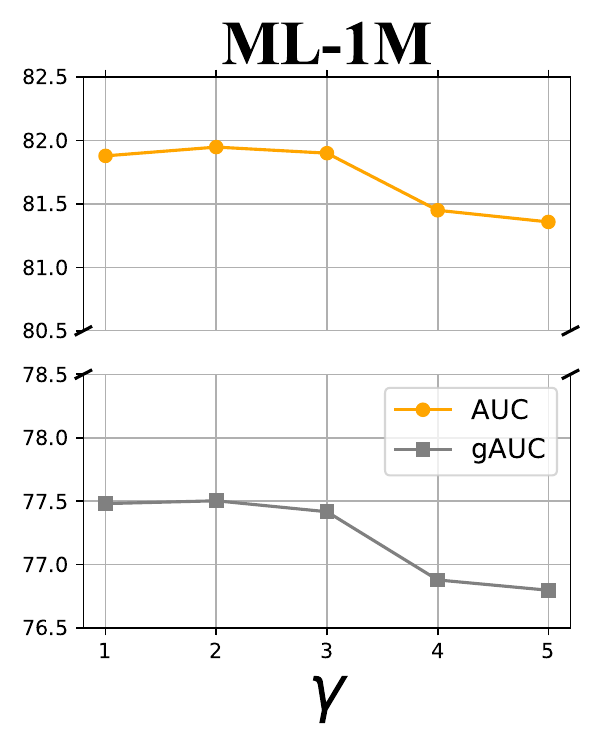}
    }
    \captionsetup{justification=raggedright}
    \caption{Influence of magnitude $\gamma$ of $\mathcal{L}_{TF}$.}
    \label{gamma}
\end{figure}

We introduce the modulating factor $\gamma$ for $\mathcal{L}_{TF}$ in order to control the strength of the rescaling loss, i.e., the larger $\gamma$ is, the more extreme the loss function is. We fixed the rest of the parameters and adjust $\gamma$ in \{1, 2, 3, 4, 5\}, and the experimental results are shown in Figure \ref{gamma}. It can be observed that as $\gamma$ increases, there is an overall trend of decreasing performance of the model, so it is not a good idea to make the losses more extreme. We think that the main reason for the decrease in performance is that as $\gamma$ increases, $\mathcal{L}_{\text{complex}}$ focuses too much on hard samples, which leads to Challenging Sample being only learned in Simple FI Encoder, hence degrades the model's performance. Meanwhile, it can be found that the model always achieves better performance when $\gamma$ is in \{1, 2, 3\}. Therefore, we propose to find a suitable $\gamma$ in this range.

\begin{figure}[t]
     \subfloat{
        \centering
        \includegraphics[width=0.32\linewidth]{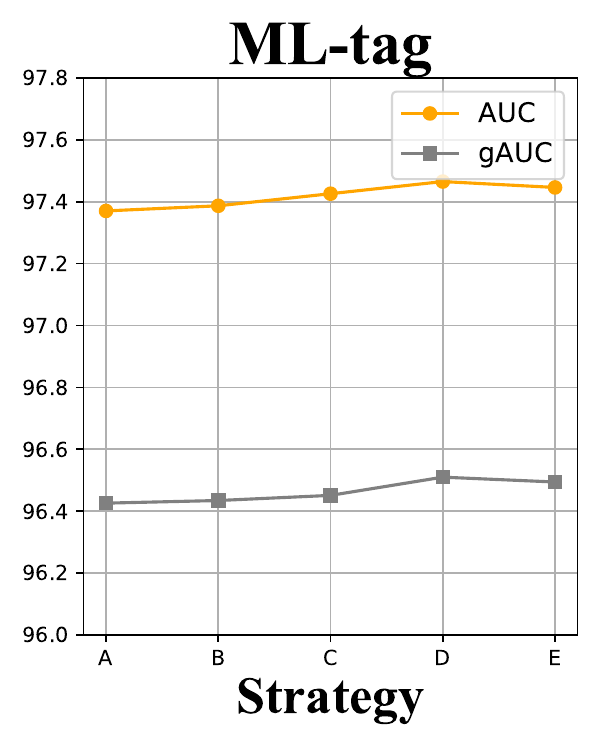}
    }
    \subfloat{
        \centering
        \includegraphics[width=0.32\linewidth]{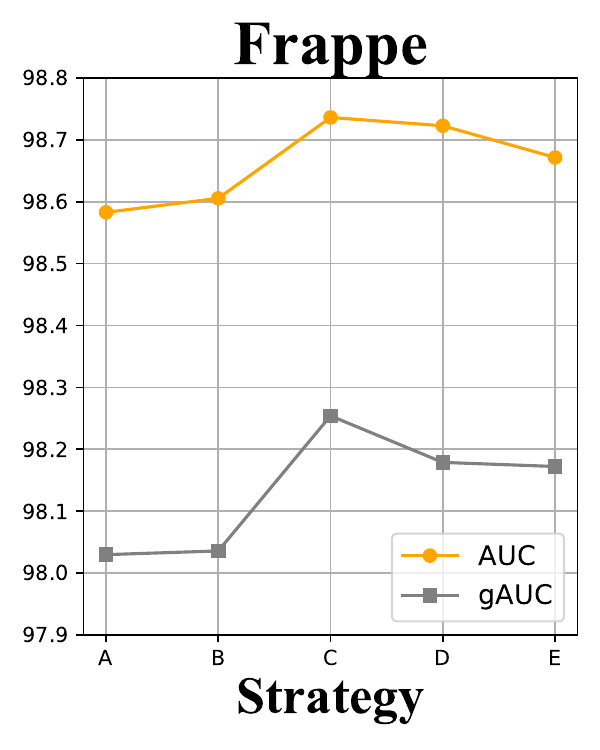}
    }
    \subfloat{
        \centering
        \includegraphics[width=0.32\linewidth]{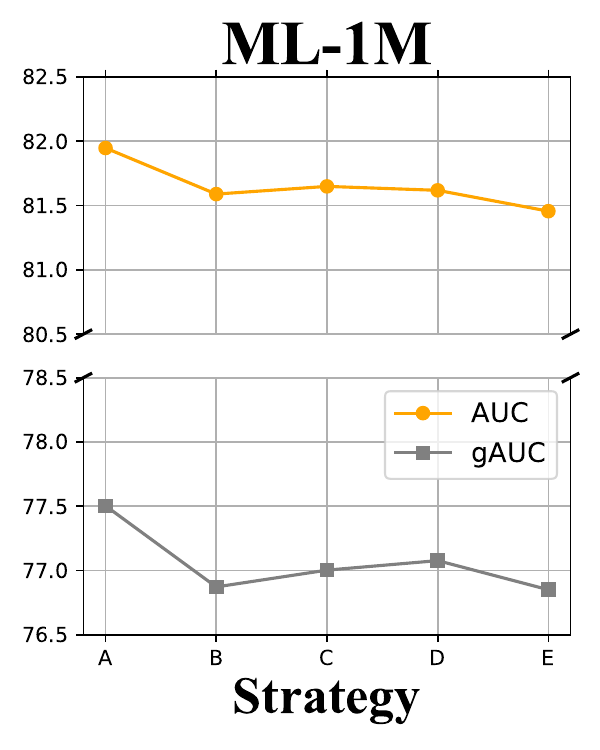}
    }
    \captionsetup{justification=raggedright}
    \caption{Influence of $\texttt{MLP}_{s}$.}
    \label{simple_encoder}
\end{figure}

\subsubsection{Impact of $\texttt{MLP}_{s}$}
To explore the simple encoder's performance across configuration, we design five structures:
\begin{itemize} 
\item  A: [400] uses a single linear layer with 400 neurons as a simple encoder.
\item  B: [800] employs a single linear layer with 800 neurons as a simple encoder.
\item  C: [200, 200] utilizes an MLP with two layers, each containing 200 neurons, as a simple encoder.
\item  D: [400, 400] uses an MLP with two layers, each containing 400 neurons, as a simple encoder.
\item  E: [800, 800] employs an MLP with two layers, each containing 800 neurons, as a simple encoder.
\end{itemize}
Experimental results, as shown in Figure \ref{simple_encoder}, indicate that the optimal structure varies due to different distributions of sample difficulty within each dataset. Structure D achieves the best performance on the ML-tag dataset, while structure C peaks on Frappe, and structure A excels on ML-1M. From these observations, we can infer that ML-tag has fewer easy samples, followed by Frappe, and ML-1M has the least challenging samples. Additionally, it is evident that having a higher number of neurons does not necessarily improve the performance of a simple encoder, despite having the most parameters, structure E does not perform well across the three datasets.

\begin{figure*}[t]
    \subfloat[Positive sample on Frappe]{
        \centering
        \includegraphics[width=0.25\linewidth]{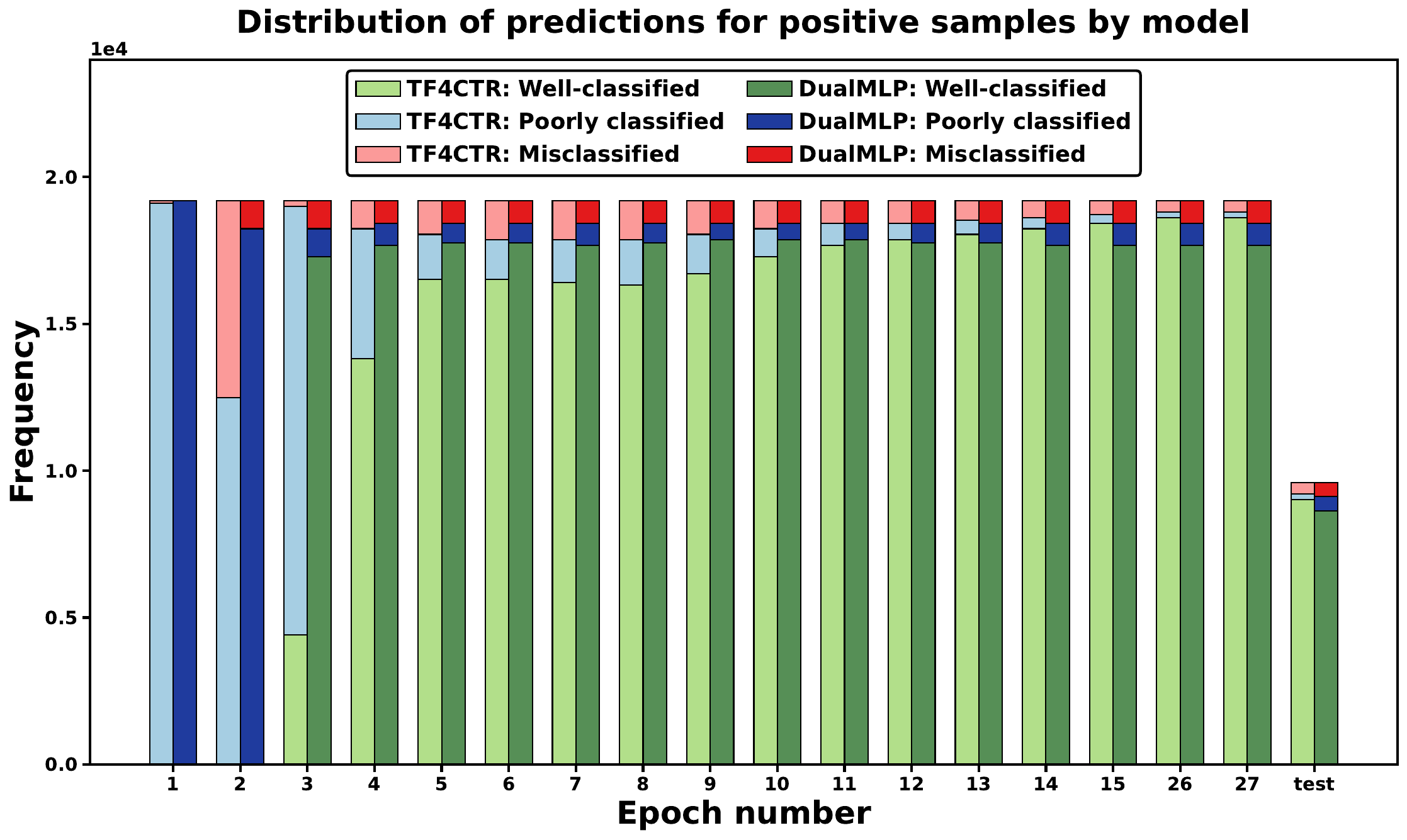}
    }
    \subfloat[Negative sample on Frappe]{
        \centering
        \includegraphics[width=0.25\linewidth]{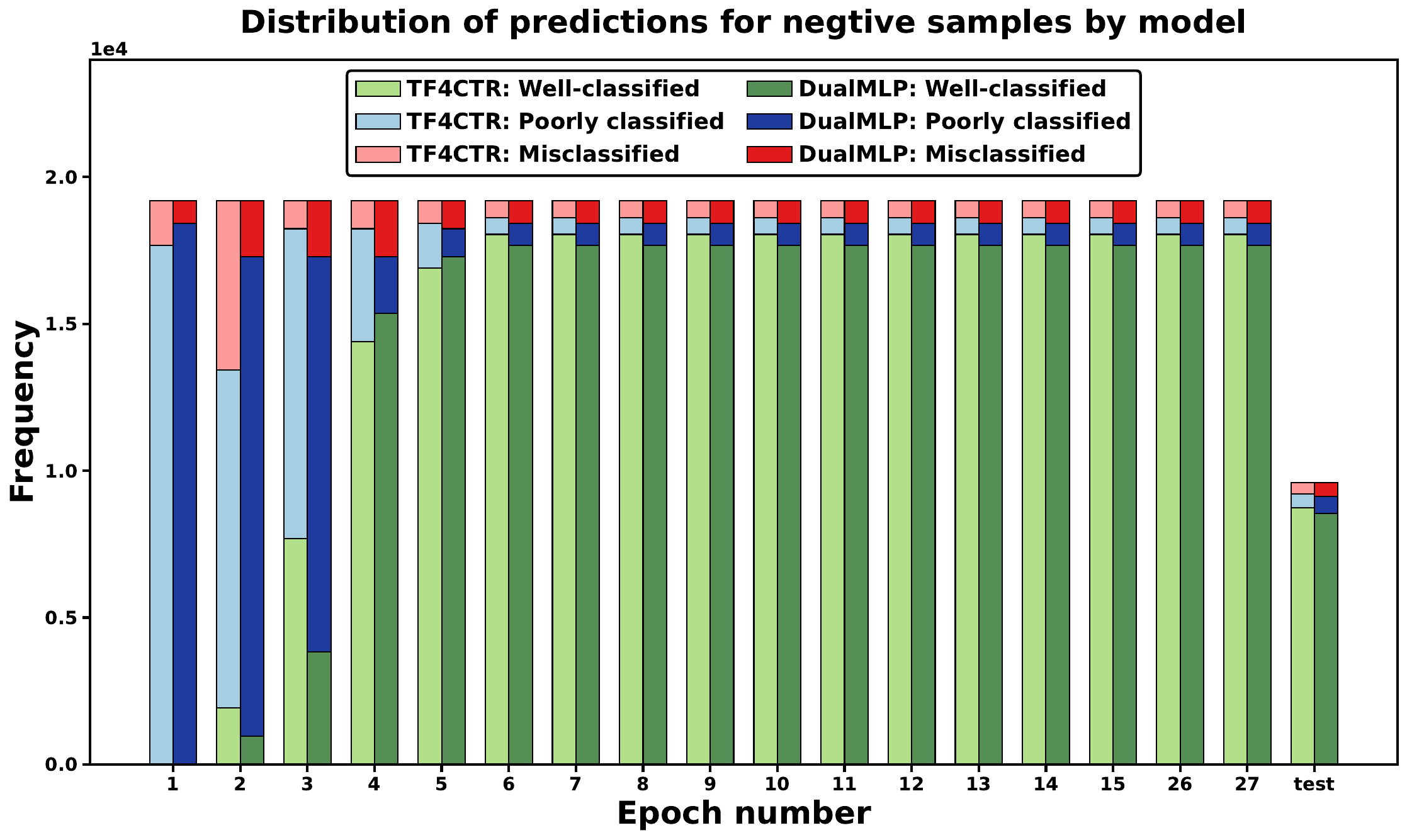}
    }
    \subfloat[Positive sample on ML-tag]{
        \centering
        \includegraphics[width=0.25\linewidth]{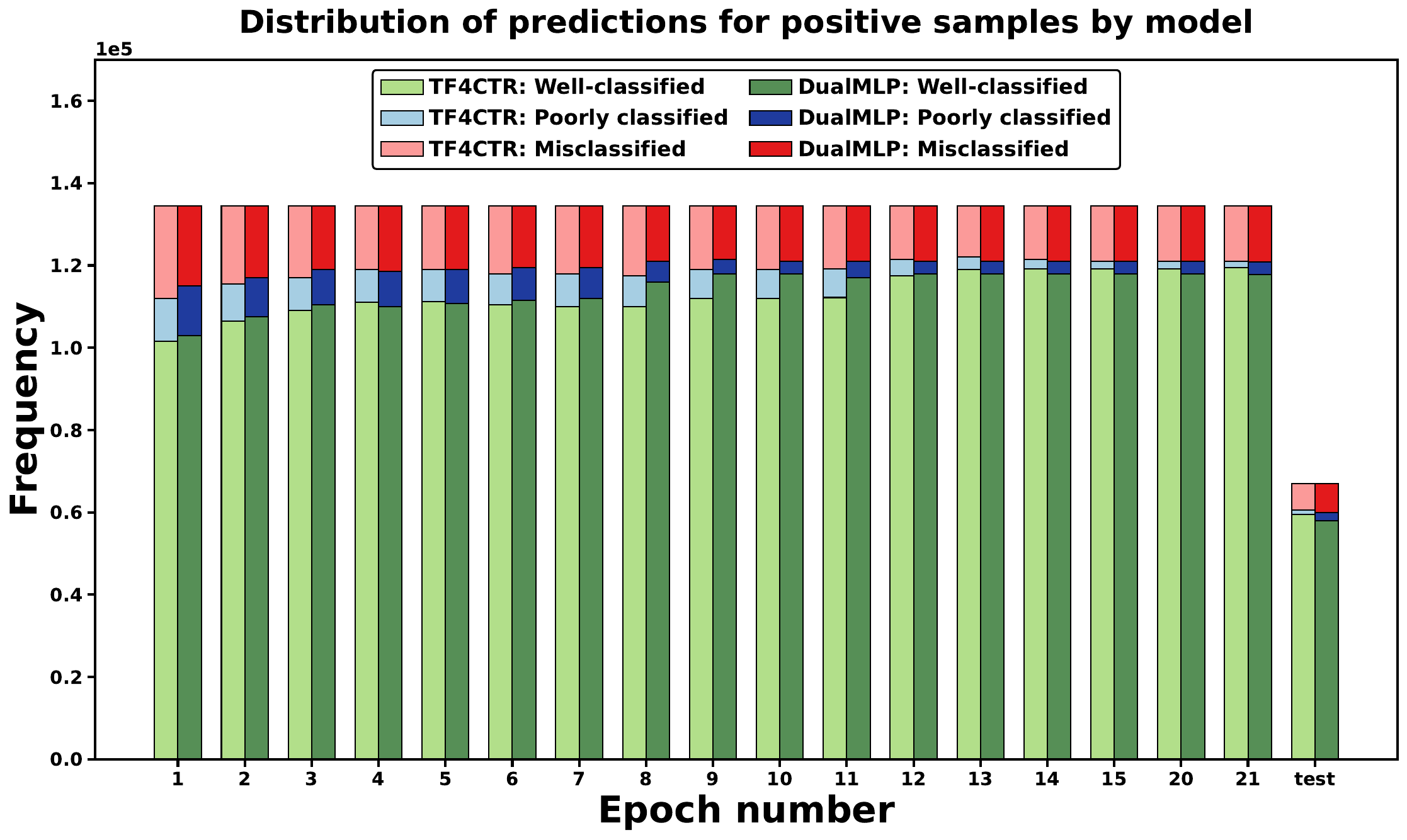}
    }
    \subfloat[Negative sample on ML-tag]{
        \centering
        \includegraphics[width=0.25\linewidth]{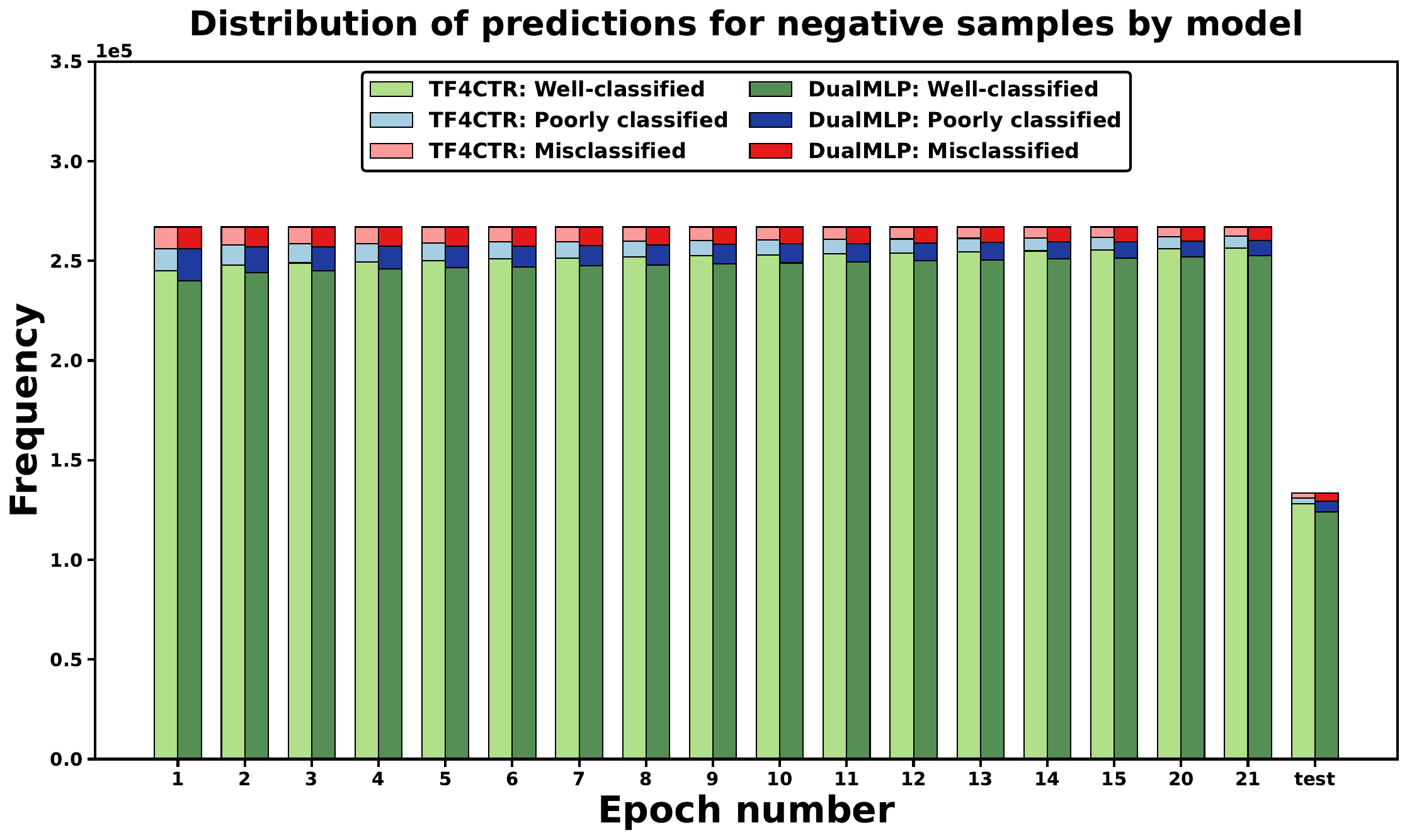}
    }
    \captionsetup{justification=raggedright}
    \caption{Visualization of the model's ability to generalize for samples of varying degrees of difficulty.}
    \label{learning}
\end{figure*}

\subsection{Model Learning Process Investigation (RQ6)}
\subsubsection{Model Generalization across Learning Stages}
To further confirm whether our method has aided DualMLP in better differentiating samples during the learning process, we adjust only the parameter \textit{early\_stop\_patience} to ensure that both models had a consistent number of training epochs. Subsequently, we categorize the prediction outcomes on validation and test sets into three types using the thresholds \{0.3, 0.6\}: \#\textit{Misclassified}, \#\textit{Poorly classified}, and \#\textit{Well-classified}. After that, These categories are visualized in a bar chart, illustrating the models' generalization capabilities on the validation set and their performance on the test set, as shown in Figure \ref{learning}.

Overall, TF4CTR has a higher count of \#\textit{Well-classified} instances than DualMLP, both on the validation and test sets. Moreover, a significant reduction in the number of \#\textit{Poorly classified} instances is clearly observed. This further validates the effectiveness of the coordinated action of SSEM, DFM, and TF Loss within the TF4CTR framework, which enables the backbone model to better distinguish challenging samples, thereby enhancing the model's performance. On the other hand, it is evident from the Frappe and ML-tag datasets that the performance of DualMLP stabilizes after the model has been trained to epoch 6 and epoch 8, respectively. In contrast, TF4CTR continues to refine its classification results, reducing the numbers of \#\textit{Misclassified} and \#\textit{Poorly classified} instances. The effectiveness of this refinement approach is further confirmed by the classification outcomes on the test set. Additionally, it is also worth mentioning that we can clearly observe that the model's ability to classify positive samples is much lower than that for negative samples, which can be attributed to the imbalance of positive and negative samples within the dataset \cite{rankandlog}.

\subsubsection{Running Time Comparison} 
\label{Running Time Comparison}
To investigate the additional training cost introduced by our proposed plug-and-play modules—SSEM and DFM—and the TF Loss, we report the training time for the first epoch and the inference time on the validation set. The experimental results depicted in Table \ref{time} are collected on an Intel(R) Xeon(R) Platinum 8336C CPU and an NVIDIA GeForce RTX 4090 GPU.

\begin{table}[htbp]
\renewcommand\arraystretch{1.2}
\tiny
\centering
\caption{Training and inference time comparison for the first epoch on KKBox. x denotes the times compared to the base model.} 
\resizebox{1 \linewidth}{!}{
\begin{tabular}{c|cc|cc}
\hline\hline
Model             & \begin{tabular}[c]{@{}c@{}}Training\\  time \end{tabular} & \begin{tabular}[c]{@{}c@{}}Relative \\ ratio\end{tabular} & \begin{tabular}[c]{@{}c@{}}Inference \\ time \end{tabular} & \begin{tabular}[c]{@{}c@{}}Relative \\ ratio\end{tabular} \\ \hline
DNN (Base)        & 23s                                                          & -                                                         & 0.11ms                                                             & -                                                         \\
xDeepFM           & 35s                                                          & 1.52x                                                     & 2.08ms                                                             & 18.91x                                                     \\
AutoInt+          & 36s                                                          & 1.56x                                                     & 0.59ms                                                             & 5.36x                                                     \\
AFN+              & 33s                                                          & 1.43x                                                     & 1.17ms                                                             & 10.64x                                                     \\
CL4CTR            & 63s                                                          & 2.73x                                                     & 0.16ms                                                             & 1.45x                                                        \\
DualMLP           & 25s                                                          & 1.08x                                                     & 0.18ms                                                             & 1.64x                                                        \\ \hline
TF4CTR (SER+MoEF) & 30s                                                          & 1.30x                                                     & 0.21ms                                                             & 1.91x                                                        \\ \hline\hline
\end{tabular}}
\label{time}
\end{table}
{
As Table~\ref{time} shows, we report the additional runtime overhead of various models in comparison to a simple DNN benchmark. It is observed that CL4CTR, due to its contrastive loss function tailored for embeddings, incurs the highest training time, yet its inference speed remains on par with the DNN. Models such as xDeepFM, AutoInt+, and AFN+ experience greater latencies in both training and inference times due to their more complex explicit encoders. With its excellent parallelization capabilities, DualMLP shows negligible difference in inference time compared to the DNN, and only a modest 1.08x increase in training time. TF4CTR and DualMLP have similar inference times, while TF4CTR introduces a moderate training-time increase over DualMLP (30s vs.\ 25s, 1.20x), and its accuracy gains are consistently observed in previous sections (Table~\ref{baseline}). In terms of inference time, the latency of TF4CTR is only 1.91 times that of DNN, slightly higher than the widely used DualMLP in industrial settings \cite{finalmlp, PEPNET}. From a performance--latency trade-off perspective, this small absolute latency increase (0.18ms $\rightarrow$ 0.21ms) comes with consistent AUC improvements across datasets (Table~\ref{baseline}). Meanwhile, as shown in the results in Table \ref{baseline}, we observe that the inference latency of TF4CTR is not only much lower than that of xDeepFM and AFN+, but its performance also outperforms them. 
Consequently, the TF4CTR's incremental training time and inference latency are acceptable, suggesting that our modules and loss function are lightweight and suitable for industrial applications.
}
\section{Related Work}
\subsection{Deep CTR Prediction}
Most existing deep CTR models can be categorized into two types. (1) User behavior sequence-based CTR prediction models \cite{DIN, DIEN, MISS, AT4CTR, TKDE1}, which focus on modeling historical sequence data at the user level, such as past clicks, browsing, or purchase records. These models understand and predict future user actions, such as the likelihood of clicking on specific ads or purchasing products. (2) Feature interaction-based CTR prediction models \cite{autohash, aim, TKDE2, EDCN, dcnv2, pnn2, masknet, fignn, GDCN}, which have broader applicability and focus on the interactions among sample features. By identifying feature combinations that significantly affect the click-through rate, these models reveal deeper data patterns. Many CTR models with parallel structures achieve excellent performance \cite{GDCN, FINAL, dcnv2}, primarily due to their ability to integrate various structured feature encoders while maintaining efficient computational performance during feature processing and learning. These models typically employ deep learning techniques to automatically learn complex interactions between features, eliminating the need for manual feature engineering, and thereby enhancing the models' generalization capabilities and predictive accuracy. For instance, attention-based models such as AFM \cite{AFM}, AutoInt \cite{autoint}, and FRNet \cite{FRNET} utilize attention mechanisms to find appropriate interaction orders among features, thereby filtering out noise and enhancing model performance. On the other hand, given the challenges MLPs face in capturing explicit low-order feature interactions \cite{neuralvsmf}, product-based models like DeepFM \cite{deepfm}, DCNv2 \cite{dcnv2}, GDCN \cite{GDCN}, DCNv3 \cite{DCNv3} leverage the interpretability and reliability of explicit feature encoders to overcome the performance limitations of MLPs. Recent studies also indicate that DualMLP's performance is not as poor as previously thought \cite{finalmlp}\cite{FINAL}, suggesting that appropriate structural adjustments to MLPs can also yield significant improvements.

\subsection{Loss Function for CTR Prediction}
CTR models aim to predict whether a user will click on a current item, thus categorizing this predictive behavior as a binary classification task. Typically, CTR prediction models employ binary cross-entropy (Log Loss) as the loss function \cite{fignn,finalmlp,dcnv2}, which optimizes each prediction outcome individually, thereby directly enhancing the accuracy of each model prediction. Meanwhile, some studies \cite{listwise,jointloss} note that CTR prediction aims to ensure positive scores exceed negative ones. Therefore, they combine pairwise rank loss and listwise loss with Log Loss to improve ranking.

With the rise of contrastive learning \cite{simGCL,XsimGCL, TOIS1, TOIS2}, contrastive loss has also been increasingly applied to CTR prediction, such as in CL4CTR \cite{CL4CTR}, which introduces concepts of feature alignment and uniformity across fields. However, the proposed contrastive module significantly increases training costs. CETN \cite{CETN} introduces concepts of diversity and homogeneity in feature representation, adaptively adjusting the learning process of multiple encoders through the synergy of Do-InfoNCE and cosine loss. In the field of computer vision, Focal Loss \cite{focalloss} initially addresses the issue of imbalanced sample difficulty by adding a modulating factor to Log Loss. However, because this factor is always less than one, it may reduce the impact of vital supervision signals. Further, R-CE Loss \cite{rce} suggests ignoring outliers to improve data credibility and help the model learn accurate feature representations.
Beyond these loss-level modifications, some works explore sample re-weighting and curriculum-style strategies (e.g., self-paced learning, OHEM, and curriculum learning) to adjust the sampling order or importance of training examples. These methods typically operate at the data or loss level under a single shared encoder. In contrast, our TF4CTR framework is designed to jointly model sample difficulty and encoder specialization: a Sample Selection Embedding Module (SSEM) routes easy, challenging, and hard samples to different encoders, while the proposed Twin Focus Loss provides encoder-specific supervision signals. This design is also orthogonal to multi-expert architectures such as MMoE~\cite{mmoe} and recent industrial CTR systems~\cite{dcnv2,DCNv3,finalmlp,FINAL}, which mainly focus on task-level heterogeneity or backbone design, whereas we focus on difficulty-aware specialization for a single CTR task.

\section{Conclusion and Future Work}
In this paper, we analyzed and identified the inherent limitations of current parallel-structured CTR prediction models, namely their lack of sample differentiation ability, unbalanced sample distribution, and undifferentiated learning process. To address these limitations, we proposed the lightweight and plug-and-play Sample Selection Embedding Module, Dynamic Fusion Module, and Twin Focus Loss to form a model-agnostic CTR framework, TF4CTR. These three additional plug-ins played roles in three positions of the feature interaction encoder's input, output, and gradient optimization, respectively, to improve the generalization ability and performance of the model. Extensive experiments on five real-world datasets demonstrated the effectiveness and compatibility of TF4CTR. In the future, we will further explore the design concept of TF4CTR and expect to introduce a loss function for challenged samples to improve the generality of the framework.

\section*{Acknowledgments}
This work is supported by the National Science Foundation of China (No. 62272001 and No. 62206002) and Hefei Key Common Technology Project (GJ2022GX15). This research was also  supported in part by the Australian Research Council Discovery Projects under Grant No. DP200102491, DP230101790, and Linkage Projects under Grant No. LP210301393, LP220100482.



 
%

\bibliographystyle{IEEEtran}
\bibliography{cite}

\begin{IEEEbiography}
[{\includegraphics[width=1in,height=1.25in,clip,keepaspectratio]{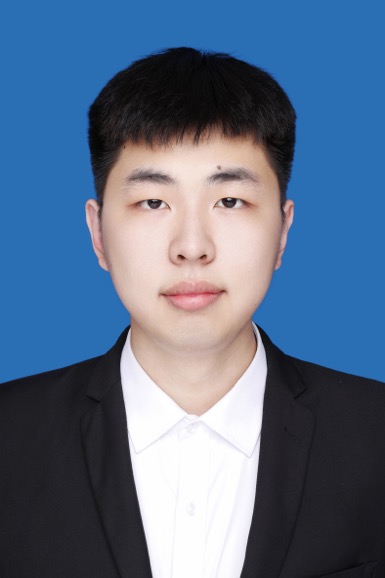}}]{Honghao Li}
received the Bachelor degree in Computer engineering and Technology from Bengbu University, Bengbu, China, in 2022. He is currently pursuing a Ph.D. degree at Anhui University's School of Computer Science and Technology. His current research interests include CTR prediction, service computing, and recommender systems.
\vspace{-1em}
\end{IEEEbiography}

\begin{IEEEbiography}
[{\includegraphics[width=1in,height=1.25in,clip,keepaspectratio]{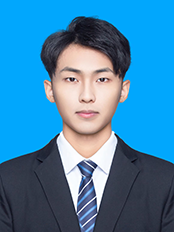}}]
{Qiuze Ru}
received the B.E. degree in Civil Engineering from Anhui Jianzhu University, Hefei, China, in 2022. He is currently pursuing the M.S. degree in Computer Science and Technology with the School of Computer Science and Technology, Anhui University. His current research interests include click-through rate prediction, recommender systems, and service computing.
\vspace{-1em}
\end{IEEEbiography}

\begin{IEEEbiography}[{\includegraphics[width=1in,height=1.25in,clip,keepaspectratio]{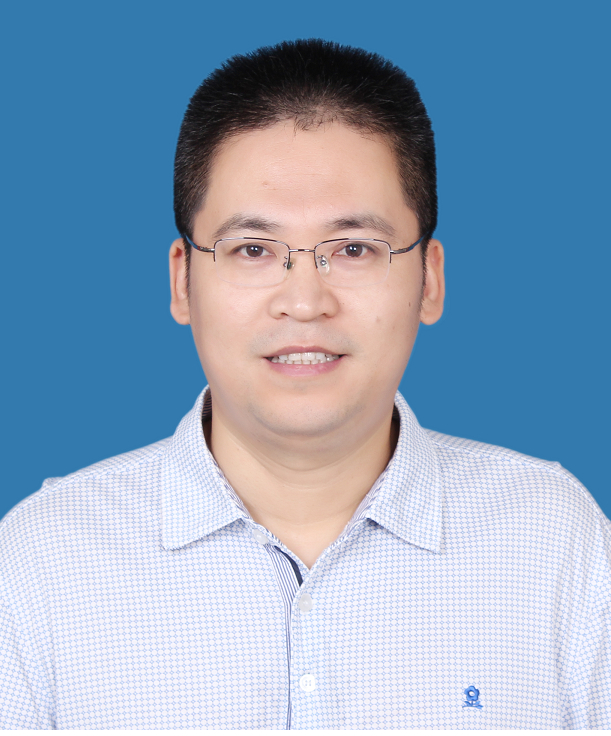}}]{Yiwen Zhang}
received the Ph.D. degree in management science and engineering from Hefei University of Technology, in 2013. He is currently a full professor with the School of Computer Science and Technology, Anhui University. He has published more than 70 papers in highly regarded conferences and journals, including IEEE TKDE, IEEE TMC, IEEE TSC, ACM TOIS, IEEE TPDS, IEEE TNNLS, ACM TKDD, SIGIR, ICSOC, ICWS, etc. His research interests include service computing, cloud computing, and big data analytics. Please see more information in our website \url{http://bigdata.ahu.edu.cn/}.
\vspace{-1em}
\end{IEEEbiography}

\begin{IEEEbiography}[{\includegraphics[width=1in,height=1.25in,clip,keepaspectratio]{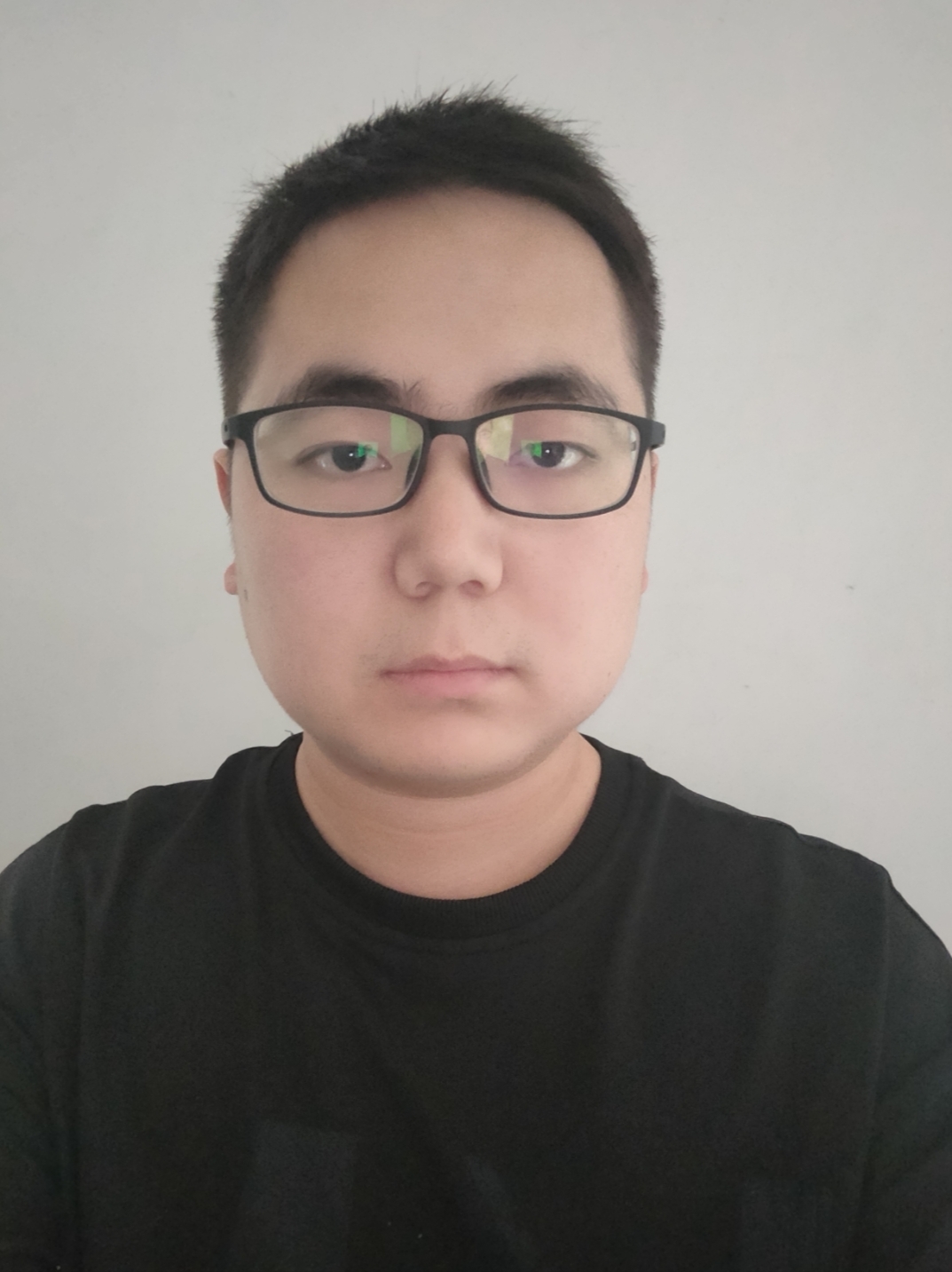}}]{Yi Zhang}
received the Bachelor  degree in Computer Science and Technology from Anhui University,
Hefei, China, in 2020, where he is currently pursuing the Ph.D. degree. He has publications in several top conferences and journals, including IEEE TKDE, IEEE TSMC, IEEE TBD, ACM TOIS, and ACM SIGIR, etc. His current research interests include graph neural network, personalized recommender systems, and service computing.
\vspace{-1em}
\end{IEEEbiography}

\begin{IEEEbiography}[{\includegraphics[width=1in,height=1.25in,clip,keepaspectratio]{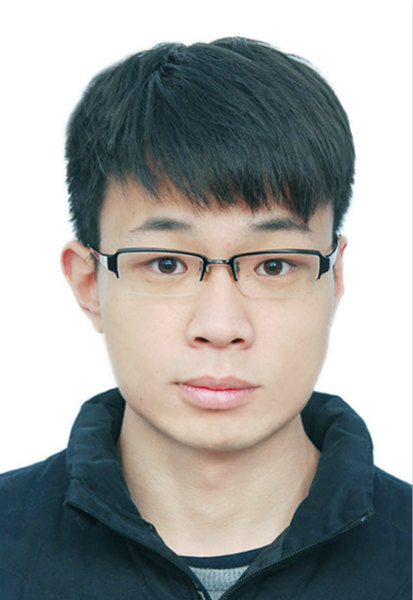}}]{Lei Sang}
received the Ph.D. degree from the Faculty of Engineering and Information Technology, University of Technology Sydney, Sydney, Australia, in 2021.
He is currently a Lecturer with the School of Computer Science and Technology, Anhui University, Anhui, China. His current research interests include natural language processing, data mining, and recommendation systems.
\vspace{-1em}
\end{IEEEbiography}

\begin{IEEEbiography}[{\includegraphics[width=1in,height=1.25in,clip,keepaspectratio]{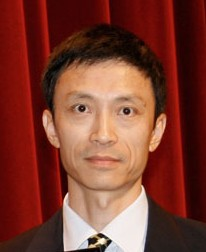}}]{Yun Yang}
(Senior Member, IEEE) received the Ph.D. degree in computer science from the University of Queensland, Australia, in 1992. He is currently a full professor with Department of Computing Technologies, Swinburne University of Technology, Melbourne, Australia. His research interests include software technologies, cloud and edge computing, workflow systems, and service computing. He was an associate editor for IEEE Transactions on Parallel and Distributed Systems during 2018-2022.
\vspace{-1em}
\end{IEEEbiography}

\end{document}